%% file: AutomaticTimestepAdjustment_arXiv.tex
\documentclass{article}
\pdfoutput=1
\usepackage[a4paper, left= 2cm, textwidth=17cm]{geometry}
\usepackage{amsmath}
\usepackage{amssymb}
\usepackage{booktabs}
\usepackage{siunitx}
\usepackage{algorithm}
\usepackage{algorithmic}
\usepackage{graphicx}
\usepackage{url}

\begin{document}

\title{Adaptive Time Step Algorithms for the Simulation of marine Ecosystem Models using the Transport Matrix Method Implementation Metos3D}

\author{Markus Pfeil, Thomas Slawig\\Kiel University, Dep. of Computer Science.\\ Correspondence: ts@informatik.uni-kiel.de}

\maketitle

\bibliographystyle{plain}

\begin{abstract}
  \input{Latexfiles/Abstract.tex}
\end{abstract}

\section{introduction}  
 \input{Latexfiles/Introduction_arxiv.tex}

\section{Model Description}
\label{sec:Model}

\input{Latexfiles/Model_arxiv.tex}

\section{Step Size Control Algorithms}
\label{sec:StepSizeControl}

\input{Latexfiles/StepSizeControl_arxiv.tex}

\section{Decreasing Time Steps Algorithm}
\label{sec:DecreasingTimesteps}

\input{Latexfiles/DecreasingTimesteps.tex}

\section{Results}
\label{sec:Results}

\input{Latexfiles/Results_arxiv.tex}

\section{Conclusions}
\label{sec:Conclusions}

\input{Latexfiles/Conclusions_arxiv.tex}

\section*{Code and data availability}
\input{Latexfiles/CodeDataAvailability_arxiv.tex}


\bibliography{AutomaticTimeStepAdjustment_arxiv.bbl}

\end{document}

%% file: Latexfiles/Abstract.tex
The reduction of the computational effort is desirable for the simulation of
marine ecosystem models. Using a marine ecosystem model, the assessment and the
validation of annual periodic solutions (i.e., steady annual cycles) against
observational data are crucial to identify biogeochemical processes, which, for
example, influence the global carbon cycle. For marine ecosystem models, the
transport matrix method (TMM) already lowers the runtime of the simulation
significantly and enables the application of larger time steps
straightforwardly. However, the selection of an appropriate time step is a
challenging compromise between accuracy and shortening the runtime. Using an
automatic time step adjustment during the computation of a steady annual cycle
with the TMM, we present in this paper different algorithms applying either an
adaptive step size control or decreasing time steps in order to use the time
step always as large as possible without any manual selection. For these methods
and a variety of marine ecosystem models of different complexity, the accuracy
of the computed steady annual cycle achieved the same accuracy as solutions
obtained with a fixed time step. Depending on the complexity of the marine
ecosystem model, the application of the methods shortened the runtime
significantly. Due to the certain overhead of the adaptive method, the
computational effort may be higher in special cases using the adaptive step size
control. The presented methods represent computational efficient methods for the
simulation of marine ecosystem models using the TMM but without any manual
selection of the time step.

%% file: Latexfiles/Introduction_arxiv.tex
In climate research, marine ecosystem models are important for the
assessment of the role of the marine ecosystem in climate change. Diverse
biogeochemical processes influence the changes of the marine ecosystem. As part
of the global carbon cycle, one aim, for example, is to identify biogeochemical
processes that affect the $\textrm{CO}_{2}$ uptake and storage of the ocean.
Marine ecosystem models take the interplay of the ocean circulation and the
biogeochemical processes into account and, therefore, consists of a global
circulation model coupled to a biogeochemical model, cf.
\cite{Fasham03, SarGru06}. The equations and variables describing
the ocean dynamics (i.e., the physical processes) are well known. In contrast,
many different biogeochemical models exist differing in the complexity by the
number of state variables and parametrizations, because there is, in general,
no set of equations and variables to describe the biogeochemical processes, see e.g.,
\cite{KrKhOs10}. Accordingly, a validation of the different
biogeochemical models is necessary including the assessment of the ability of
the model output to reproduce observational data, cf.
\cite{FLSW01, SchOsc03}. This assessment involves both a parameter
optimization and a discussion of the simulation results.

Simulation runs with marine ecosystem models are computationally expensive. In
three spatial dimensions, the simultaneous computation of the ocean circulation
and the biogeochemical model already results in a high computational effort for
a single evaluation of a marine ecosystem model \cite{Osc06}. Thus, the
computational effort of the computation of a steady annual cycle is enormous
because it requires a long-time integration over several millennia, cf.
\cite{BeDiWu08, Bryan84, DaMcLa96, WunHei08, SibWun11}. For applications
requiring a high number of model evaluations (such as parameter sensitivity, e.g.,
uncertainty or identification studies \cite{Kri17, KSKSO17}), the high
computational effort becomes even more obstructive.

Several strategies focus on the reduction of the computational effort to
compute a steady annual cycle of a marine ecosystem model, e.g.,
\cite{Bryan84, DaMcLa96, Wang01, KhViCa05}. Using domain decomposition
methods, parallelization, firstly, reduces the runtime while introducing some additional computational effort and data transfer. Instead of
using a fully coupled simulation (also called \emph{online} simulation), the 
\emph{offline} model, secondly, reduces the computational effort. More
specifically, an offline model considers only the one-way coupling of the
influence of the ocean circulation on the biogeochemical model. Due to the
neglect of the impact of the biogeochemical model on the circulation, an offline
model can use pre-computed data of the ocean circulation. \cite{KhViCa05},
thirdly, introduced with the \emph{transport matrix method} (TMM) a method
reducing the computational effort with a tolerable loss of accuracy, see also
\cite{Kha07}. The TMM approximates the computation of the global
ocean circulation by matrix-vector multiplications and, hence, decouples the
evaluation of the biogeochemical model from the ocean circulation. In the work we present here, these three 
measures to reduce the runtime were used. However, the algorithms for step size adjustment introduced here  are not limited to them. Moreover,
\cite{Kha08} replaced the long-time integration in the TMM by the use of
Newton's method. Previous experiences with Newton's method in \cite{PiwSla16} showed that the convergence of the method, at least for more complex models, strongly depends on the initial value. This is a reason why we concentrated on the spin-up in this paper and did not apply the step size control in Newton's method, which would be nevertheless possible. Lastly, the computation of steady annual cycles using graphics
processing units shortens the computational time \cite{SiPiSl13}. Also there, a step size adjustment is possible.

The used time step affects both the computational effort and the accuracy of
the steady annual cycle computation. The application of larger time steps
obviously reduces the computational effort whereas the accuracy of the steady
annual cycle approximation decreases with a larger time step. In many cases and,
specifically, for less complex models, approximations of the same steady annual
cycles as the ones obtained with the standard time step can be computed with
bigger ones \cite{PfeSla21a}. The obtained steady annual cycles were the same
as for small time steps, but achieved with slightly reduced accuracy. In fact,
some studies using the TMM for parameter optimization, where the runtime is a
crucial point, have already used larger time steps, e.g., \cite{PPKOS13,
KSKSO17}. However, the selection of a suitable time step is a challenging task,
also in the TMM. The time step should be chosen as large as possible to keep the
computational effort low, but the accuracy of the approximation cannot be
neglected. In this paper, we present methods adjusting automatically the time
steps during the computation of the steady annual cycle. One approach is an
adaptive step size control that, depending on an error estimation, adjusts the
time step during the simulation. Here, we studied two variants: One ignores any negative tracer 
concentration that might occur during the spin-up. The other one 
detects  negative values and includes this in the step size control, i.e., if negative values 
occur for an increased time step, the bigger step size is not accepted. 
 The third algorithm we used starts with a possible large time step and uses
it as long as possible before the time step is decreased. For this purpose, the
method checks the progress of the steady annual cycle computation using the
current time step. In particular, all methods use the largest possible time step
to shorten the runtime of the simulation of marine ecosystem models.

This paper is structured as follows: Sect. \ref{sec:Model} contains a
description of marine ecosystem models including the computation of steady
annual cycles. In Sects. \ref{sec:StepSizeControl} and
\ref{sec:DecreasingTimesteps}, we introduce the three methods used to compute a
steady annual cycle with an automatic time step adjustment. Numerical results
for the steady annual cycle computation are presented in Sect. \ref{sec:Results}.
The paper closes with a summary and conclusions (Sect. \ref{sec:Conclusions}).

%% file: Latexfiles/Model_arxiv.tex
A marine ecosystem model represents the interaction between the ocean
circulation, the ocean biota and marine biogeochemical cycles. This involves
modeling the marine ecosystem by a given number of ecosystem species (or
biogeochemical tracers), which are substances in the ocean water and subject to
chemical or biochemical reactions. Due to the full coupling of the ocean
circulation with the tracers, in which the circulation influences the tracer
concentrations and, vice versa, the tracer concentrations affect the
circulation, the simulation of a fully coupled model (\emph{online} model) is
computationally expensive. Restricting
this coupling to the influence of the ocean circulation on the tracer
concentrations (i.e., an \emph{offline} model using \emph{passive tracers} that
do not affect the ocean circulation) reduces the computational effort. In
particular, this one-way coupling enables the application of a pre-computed
ocean circulation for the simulation.

\subsection{Model Equations for marine Ecosystems}
\label{sec:ModelEquations}

  A system of partial differential equations describes the marine ecosystem
  model. The number of modeled tracers defines the complexity of the marine
  ecosystem model and, hence, the size of the system of differential equations.
  For $n_y \in \mathbb{N}$ tracers on a spatial domain $\Omega \subset
  \mathbb{R}^3$ (i.e., the whole global ocean) and a time interval $[0,1]$ (i.e., one model
  year), we consider marine ecosystem models using an offline
  model. With the function $y_i: \Omega \times [0,1] \rightarrow \mathbb{R}$,
  $i \in \left\{1, \ldots, n_y \right\}$, of the tracer concentrations for the
  single tracer $y_i$ and the vector $\vec{y} := \left( y_i
  \right)_{i=1}^{n_{y}}$ of all tracers, the system of parabolic partial
  differential equations
  \begin{align}
  \label{eqn:Modelequation}
      \frac{\partial y_i}{\partial t} (x,t)
           + \left( D (x,t) + A(x,t) \right) y_i (x,t)
        &= q_i \left( x, t, \vec{y}, \vec{u} \right),\quad 
         x \in \Omega, t \in [0,1], \\
    \label{eqn:Boundarycondition}
      \frac{\partial y_i}{\partial n} (x,t) &= 0,
        \quad x \in \partial \Omega, t \in [0,1],
  \end{align}
  $i \in\{ 1, \ldots, n_{y}\}$, represents the tracer transport of a marine
  ecosystem model. 
  In Eq. \eqref{eqn:Modelequation}, $D$ denotes the diffusion, $A$ the advection operator, and $q$ represents the biogeochemical model describing the interaction between the different tracers.
  The homogeneous Neumann boundary conditions
  \eqref{eqn:Boundarycondition} including the normal derivative model that there are no
  fluxes on the boundary.

  The advection and diffusion, coming from the ocean circulation, determines the
  tracer transport in the ocean. For a given velocity field $v: \Omega
  \times [0,1] \rightarrow \mathbb{R}^3$, the linear operator $A: \Omega \times
  [0,1] \rightarrow \mathbb{R}$ models the advection as
  \begin{align}
    \label{eqn:Advection}
    A(x,t) y_i (x,t) &:= \textrm{div} \left( v(x,t) y_i (x,t) \right),
    \quad x \in \Omega, t \in [0,1],
  \end{align}
  for $i \in \{1, \ldots, n_y\}$. The diffusion $D: \Omega \times [0,1]
  \rightarrow \mathbb{R}$ used as model of the turbulent effects of the ocean
  circulation neglects the molecular diffusion of the tracers themselves since
  this is known to be much smaller in relation to the diffusion caused by
  turbulence. As a result of the quite different scales in horizontal and
  vertical direction, the diffusion operator is split into a sum $D = D_h +
  D_v$ of a horizontal and vertical part in order to treat the vertical part
  in the time integration implicitly. The diffusion is split into horizontal and vertical part as
  \begin{align}
    \label{eqn:Diffusion-horizontal}
    D_{h} (x,t) y_i (x,t) &:= -{\textrm{div}_h} \left( \kappa_h (x,t) \nabla_h
                                y_i (x,t) \right),
     \quad x \in \Omega, t \in [0,1],\\
    \label{eqn:Diffusion-vertical}
    D_{v} (x,t) y_i (x,t) &:= -\frac{\partial}{\partial z}
            \left(\kappa_{v} (x,t)\frac{\partial y_i}{\partial z}(x,t)\right),
    \quad x \in \Omega, t \in [0,1],
  \end{align}
  for $i \in \{1, \ldots, n_y\}$. Here, $\textrm{div}_h$ and $\nabla_h$ denote
  the horizontal divergence and gradient, respectively, and $z$ the vertical
  coordinate. The diffusion coefficient fields $\kappa_h, \kappa_v: \Omega
  \times [0,1] \rightarrow \mathbb{R}$ are the same for all tracers due to the
  fact that the molecular diffusion is generally assumed to be smaller than the
  diffusion induced by the turbulence of the ocean circulation.

  The biogeochemical model contains the biogeochemical processes within the
  marine ecosystem. The interplay of the biogeochemical model with the effects
  of the ocean circulation (i.e., the whole system \eqref{eqn:Modelequation} to
  \eqref{eqn:PeriodicCondition}), on the other hand, is called the marine ecosystem
  model. For the tracer $y_i$, $i \in \{1, \ldots, n_y\}$, the, in general,
  nonlinear function $q_i: \Omega \times [0,1] \rightarrow~\mathbb{R}, \left(
  x, t \right) \mapsto q_i \left( x, t, \vec{y}, \vec{u} \right)$
  describes the biogeochemical processes for this tracer. In particular, this
  nonlinear function $q_i$ includes firstly the influence of the variability of
  the solar radiation at space and time, secondly the coupling of this tracer
  to the other species and thirdly $n_u \in \mathbb{N}$ model parameters
  $\vec{u} \in \mathbb{R}^{n_u}$ controlling, for example, growth, loss and
  mortality rates or sinking speed of this tracer. Altogether, the
  biogeochemical model $\vec{q} = \left( q_i \right)_{i=1}^{n_{y}}$
  summarizes the biogeochemical processes of all tracers.

  An annual periodic solution of the marine ecosystem (i.e., a steady annual
  cycle) satisfies in addition to \eqref{eqn:Modelequation} and 
  \eqref{eqn:Boundarycondition}
  \begin{align}
   \label{eqn:PeriodicCondition}
    y_i (x, 0) &= y_i (x, 1), \quad x \in \Omega,
  \end{align}
  for $i \in \{1, \ldots, n_y\}$. For this purpose, we assume that the
  operators $D, A$ and the functions $q_i$ are also annually periodic in time.

\subsection{Semi-discrete Setting}
  For the adaptive step size control, we used a semi-discrete setting where
  the computational domain is already discretized in space, but time is kept
  continuous. Then  Eq. \eqref{eqn:Modelequation}, \eqref{eqn:Boundarycondition}
  read
  \begin{align}
    \frac{\partial \vec{y}_i}{\partial t} (t) + 
       \left( \mathbf{D}(t) + \mathbf{A}(t) \right) \vec{y}_{i}(t) 
    &= \vec{q}_{i}(t,\vec{y}(t), \vec{u}), 
    \quad t \in [0,1],
  \end{align}
  with initial value $\vec{y}_{i}(0)=\vec{y}_{i}^{0}$.

\subsection{Biogeochemical Models}
\label{sec:BiogeochemicalModels}

  The biogeochemical models, used in this paper, differed in the number of
  ecosystem species. \cite{KrKhOs10} introduced a hierarchy of five
  biogeochemical models with an increasing complexity starting with a simple
  model including only one tracer up to a model with five tracers. In addition
  to this hierarchy, we applied the biogeochemical model of
  \cite{DuSoScSt05}. In the following, we briefly describe the
  biogeochemical models and refer the reader to \cite{KrKhOs10, PiwSla16, DuSoScSt05}
  for a detailed description of the modeled processes and model equations. Table
  \ref{table:Parameter-Modelhierarchy} summarizes the model parameters of the
  different biogeochemical models.

  \begin{table*}[tb]
    \caption{Model parameters of the biogeochemical models.}
    \label{table:Parameter-Modelhierarchy}
    \centering
    \begin{tabular}{l l l}
      \hline
      Parameter & Description & Unit \\
      \hline
      $k_w$ & Attenuation coefficient of water & \si{m^{-1}} \\
      $k_c$ & Attenuation coefficient of phytoplankton & \si{(mmol\, P\, m^{-3})^{-1} m^{-1}} \\
      $\mu_P$ & Maximum growth rate & \si{d^{-1}} \\
      $\mu_Z$ & Maximum grazing rate & \si{d^{-1}} \\
      $K_N$ & Half saturation constant for \si{PO_4} uptake & \si{mmol\, P\, m^{-3}} \\
      $K_P$ & Half saturation constant for grazing & \si{mmol\, P\, m^{-3}} \\
      $K_I$ & Light intensity compensation & \si{W\, m^{-2}} \\
      $\sigma_Z$ & Fraction of production remaining in \si{Z} & \si{1} \\
      $\sigma_\text{DOP}$ & Fraction of phytoplankton and zooplankton losses assigned to \si{DOP} & \si{1} \\
      $\lambda_P$ & Linear phytoplankton loss rate & \si{d^{-1}} \\
      $\kappa_P$ & Quadratic phytoplankton loss rate & \si{(mmol\, P\, m^{-3})^{-1} d^{-1}} \\
      $\lambda_Z$ & Linear zooplankton loss rate & \si{d^{-1}} \\
      $\kappa_Z$ & Quadratic zooplankton loss rate & \si{(mmol\, P\, m^{-3})^{-1} d^{-1}} \\
      $k_c$ & Attenuation coefficient of phytoplankton & \si{(mmol\, P\, m^{-3})^{-1} d^{-1}} \\
      $\lambda'_P$ & Phytoplankton mortality rate & \si{d^{-1}} \\
      $\lambda'_Z$ & Zooplankton mortality rate & \si{d^{-1}} \\
      $\lambda'_D$ & Degradation rate & \si{d^{-1}} \\
      $\lambda'_\text{DOP}$ & Decay rate & \si{yr^{-1}} \\
      $b$ & Implicit representation of sinking speed & \si{1}  \\
      $a_D$ & Increase of sinking speed with depth & \si{d^{-1}} \\
      $b_D$ & Initial sinking speed & \si{m\,d^{-1}} \\
      \hline
    \end{tabular}
  \end{table*}

  \begin{table*}[tb]
    \caption{Reference parameter values of the biogeochemical models taken from
             \cite{KrKhOs10} as well as lower $(\vec{b}_\ell)$ and upper
             $(\vec{b}_u)$ bounds for the parameter values used to generate the
             Latin hypercube sample.}
    \label{table:ParameterValues-Modelhierarchy}
    \centering
    \begin{tabular}{l r r r r r c c}
      \hline
      Parameter & N & N-DOP & NP-DOP & NPZ-DOP & NPZD-DOP & $\vec{b}_\ell$ & $ \vec{b}_u$ \\
      \hline
      $k_w$         & 0.02  & 0.02  & 0.02  & 0.02  & 0.02  & 0.01  & 0.05 \\
      $k_c$         &       &       & 0.48  & 0.48  & 0.48  & 0.24  & 0.72 \\
      $\mu_P$       & 2.0   & 2.0   & 2.0   & 2.0   & 2.0   & 1.0   & 4.0 \\
      $\mu_Z$       &       &       & 2.0   & 2.0   & 2.0   & 1.0   & 4.0 \\
      $K_N$         & 0.5   & 0.5   & 0.5   & 0.5   & 0.5   & 0.25  & 1.0 \\
      $K_P$         &       &       & 0.088 & 0.088 & 0.088 & 0.044 & 0.176 \\
      $K_I$         & 30.0  & 30.0  & 30.0  & 30.0  & 30.0  & 15.0  & 60.0 \\
      $\sigma_Z$    &       &       &       & 0.75  & 0.75  & 0.05  & 0.95 \\
      $\sigma_\text{DOP}$  & & 0.67 & 0.67  & 0.67  & 0.67  & 0.05  & 0.95 \\
      $\lambda_P$   &       &       & 0.04  & 0.04  & 0.04  & 0.02  & 0.08 \\
      $\kappa_P$    &       &       & 4.0   &       &       & 2.0   & 6.0 \\
      $\lambda_Z$   &       &       &       & 0.03  & 0.03  & 0.015 & 0.045 \\
      $\kappa_Z$    &       &       &       & 3.2   & 3.2   & 1.6   & 4.8 \\
      $\lambda'_P$  &       &       & 0.01  & 0.01  & 0.01  & 0.005 & 0.015 \\
      $\lambda'_Z$  &       &       &       & 0.01  & 0.01  & 0.005 & 0.015 \\
      $\lambda'_D$  &       &       &       &       & 0.05  & 0.025 & 0.1 \\
      $\lambda'_\text{DOP}$ & & 0.5 & 0.5   & 0.5   & 0.5   & 0.25  & 1.0 \\
      $b$           & 0.858 & 0.858 & 0.858 & 0.858 &       & 0.7   & 1.5 \\
      $a_D$         &       &       &       &       & 0.058 & 0.029 & 0.087 \\
      $b_D$         &       &       &       &       & 0.0   & 0.0   & 0.0 \\
      \hline
    \end{tabular}
  \end{table*}

  Especially, the biological production depends on the available light. The
  light intensity decreases with the depth wherefore the ocean is divided into
  two layers, a euphotic (sun lit) layer of about \si{100 m} and an
  aphotic zone below. Depending on the insolation based on the astronomical
  formula of \cite{PalPla76}, the light limitation function $I: \Omega
  \times [0, 1] \rightarrow \mathbb{R}_{\geq 0}$ models the available light
  taking the ice cover and the exponential attenuation of water into account.
  Since the main part of the biological production occurs in the euphotic
  layer, particulate matter sinks from the euphotic layer to depth and
  remineralizes there according to the empirical power-law relationship
  \cite{MKKB87}.

  The N model is the simplest biogeochemical model of the hierarchy and
  represents only phosphate ($\textrm{PO}_4$) as inorganic nutrients, cf. 
  \cite{BacMai90, KrKhOs10}. The available nutrients and light
  restrict the phytoplankton production (or biological uptake). The
  phytoplankton production
  \begin{align}
    \label{eqn:Phytoplankton}
    f_P: \Omega \times [0,1] \rightarrow \mathbb{R},
      f_P (x, t) &= \mu_P y_P^* \frac{I(x,t)}{K_I + I(x,t)}
                    \frac{\vec{y}_N (x,t)}{K_N + \vec{y}_N (x,t)}
  \end{align}
  depends on the maximum production rate $\mu_P$ and applies an implicitly
  prescribed concentration of phytoplankton $y_P^* =
  0.0028$~\si{mmol\, P\, m^{-3}}. Altogether, $n_u = 5$ model parameters
  listed in Table \ref{table:ParameterValues-Modelhierarchy} control the
  biogeochemical processes of the nutrient tracer $\vec{y} =
  (\vec{y}_{\textrm{N}})$.

  The N-DOP model includes dissolved organic phosphorus (\textrm{DOP}) in
  addition to nutrients (\textrm{N}), i.e., $\vec{y} = (\vec{y}_{\textrm{N}},
  \vec{y}_{\textrm{DOP}})$, cf. \cite{BacMai91, PaFoBo05, KrKhOs10}. This
  model computes the phytoplankton production also with
  \eqref{eqn:Phytoplankton} and introduces $n_u = 7$ model parameters
  (Table \ref{table:ParameterValues-Modelhierarchy}).

  The NP-DOP model contains phytoplankton (\textrm{P}) in addition to nutrients
  (\textrm{N}) and dissolved organic phosphorus (\textrm{DOP}), i.e., $\vec{y} =
  (\vec{y}_{\textrm{N}}, \vec{y}_{\textrm{P}}, \vec{y}_{\textrm{DOP}})$, cf.
  \cite{KrKhOs10}. As a result of the explicit treatment of
  phytoplankton, the NP-DOP model computes the phytoplankton production again
  with \eqref{eqn:Phytoplankton} but using the explicit phytoplankton
  concentration $\vec{y}_{\textrm{P}}$ instead of $y_P^*$. A quadratic loss term
  for phytoplankton, moreover, models the zooplankton grazing
  \begin{align}
    \label{eqn:Zooplankton}
    f_Z: \Omega \times [0,1] \rightarrow \mathbb{R},
    f_Z (x,t) &= \mu_Z y_Z^* \frac{\vec{y}_P(x,t)^2}{K_P^2 + \vec{y}_P(x,t)^2}
  \end{align}
  with the implicitly prescribed zooplankton concentration
  $y_Z^* = 0.01$~\si{mmol\, P\, m^{-3}}. The $n_u = 13$ model parameters
  listed in Table \ref{table:ParameterValues-Modelhierarchy} control the
  biogeochemical processes of the NP-DOP model.

  The NPZ-DOP model consists of four tracers, nutrients (\textrm{N}),
  phytoplankton (\textrm{P}), zooplankton (\textrm{Z}) and dissolved organic
  phosphorus (\textrm{DOP}), i.e., $\vec{y} = (\vec{y}_{\textrm{N}},
  \vec{y}_{\textrm{P}}, \vec{y}_{\textrm{Z}},
  \vec{y}_{\textrm{DOP}})$, cf. \cite {KrKhOs10}. While the phytoplankton
  production \eqref{eqn:Phytoplankton} is the same as for the NP-DOP model, the
  zooplankton grazing \eqref{eqn:Zooplankton} explicitly contains the
  zooplankton concentration $\vec{y}_Z$ instead of the implicitly
  prescribed concentration $y_Z^*$. The model introduces $n_u = 16$ model
  parameters summarized in Table \ref{table:ParameterValues-Modelhierarchy}.

  The NPZD-DOP model, the most complex model of the hierarchy, finally,
  contains detritus (\textrm{D}) in addition to nutrients (\textrm{N}),
  phytoplankton (\textrm{P}), zooplankton (\textrm{Z}) and dissolved organic
  phosphorus (\textrm{DOP}), i.e., $\vec{y} = ( \vec{y}_{\textrm{N}},
  \vec{y}_{\textrm{P}}, \vec{y}_{\textrm{Z}}, \vec{y}_{\textrm{D}},
  \vec{y}_{\textrm{DOP}})$, cf. \cite{SOGES05, KrKhOs10}. Both the
  phytoplankton production \eqref{eqn:Phytoplankton} and the zooplankton grazing
  \eqref{eqn:Zooplankton} are identical to the NPZ-DOP model. Table
  \ref{table:ParameterValues-Modelhierarchy} lists the $n_u = 18$ model
  parameters of the NPZD-DOP model.

  The MITgcm-PO4-DOP model contains phosphate ($\textrm{PO}_4$) and dissolved
  organic phosphorus (\textrm{DOP}). This model introduced by
  \cite{DuSoScSt05} resembles the N-DOP
  model. It also has $n_u = 7$ model
  parameters which we identified with the models parameters of the N-DOP model
  (Table \ref{table:ParameterValues-Modelhierarchy}).

\subsection{Transport Matrix Method}
\label{sec:TMM}

  The \emph{transport matrix method} (TMM) reduces the simulation of the tracer
  transport of an offline model to matrix-vector multiplications.
  \cite{KhViCa05} applied a linear matrix equation instead of directly
  implementing a discretization scheme for the advection-diffusion equation
  \eqref{eqn:Modelequation} because the application of the operators A and D for
  the advection and diffusion on a spatially discretized tracer vector is linear, see also
  \cite{Kha07}. Consequently, the TMM approximates the ocean
  circulation by matrices which include the influence of all parameterized
  processes represented in the underlying ocean circulation model on the
  transport.

  Each time step of the simulation with the TMM requires only two matrix-vector
  multiplications and an evaluation of the biogeochemical model. For the
  discretization of the advection-diffusion equation, let $\left( x_k
  \right)_{k=1}^{n_x}$ a spatial discretization with $n_x \in \mathbb{N}$ grid
  points of the domain $\Omega$ (i.e., the ocean) and the time steps $t_0,
  \ldots, t_{n_{t}} \in [0,1]$, $n_t \in \mathbb{N}$, specified by
  \begin{align}
    t_j &:= j \Delta t, \quad j = 0, \ldots, n_t, \quad \Delta t := \frac{1}{n_t},
  \end{align}
  an equidistant grid of the time interval $[0,1]$ (i.e., one model year). For
  a time instant $t_j$, $j \in \{0, \ldots, n_{t} - 1\}$, the vector
  $\vec{y}_{ji} \approx \left( y_{i} \left( t_{j}, x_{k} \right)
  \right)_{k=1}^{n_x} \in \mathbb{R}^{n_x}$ is a spatial discretization of the
  tracer $y_i$, $i \in \{1, \ldots, n_y\}$, and $\vec{q}_{ji} \approx \left(
  q_i \left( x_k, t_j, \vec{y_j}, \vec{u} \right) \right)_{k=1}^{n_x} \in
  \mathbb{R}^{n_x}$ the spatially discretized biogeochemical term $q_i$ for the
  tracer $y_i$. Besides, $\vec{y}_{j} := \left( \vec{y}_{ji}
  \right)_{i=1}^{n_y} \in \mathbb{R}^{n_y n_x}$ and $\vec{q}_j := \left(
  \vec{q}_{ji} \right)_{i=1}^{n_y} \in \mathbb{R}^{n_y n_x}$ summarize the
  tracer discretization as well as the biogeochemical terms for all tracers
  using a reasonable concatenation. Discretizing the advection and horizontal
  diffusion explicitly and the vertical diffusion implicitly, the application
  of a semi-discrete Euler scheme for \eqref{eqn:Modelequation} yields a
  time-stepping
  \begin{align}
  \label{eqn:tmmtimestep}
    \vec{y}_{j+1} &=  \left( \mathbf{I} + \Delta t \mathbf{A}_j
                            + \Delta t \mathbf{D}_j^h \right) \vec{y}_j
                     + \Delta t \mathbf{D}_j^v \vec{y}_{j+1}
                     + \Delta t \vec{q}_j \left( \vec{y}_j, \vec{u} \right),
                     \quad j = 0, \ldots, n_t -1,
  \end{align}
  with the identity matrix $\mathbf{I} \in \mathbb{R}^{n_x \times n_x}$ and the
  spatially discretized counterparts $\mathbf{A}_j, \mathbf{D}_j^h$ and
  $\mathbf{D}_j^v$ of the operators $A, D_h$ and $D_v$ at time instant $t_j$,
  $j \in \{0, \ldots, n_t - 1\}$. The matrices $\mathbf{D}_{j}^{v}$ are
  block-diagonal since they involve only the vertical part of the diffusion.
  Thus, each water column is separated from the others. We denote the explicit
  and implicit transport matrices by
  \begin{align}
    \label{eqn:tmexp}
    \mathbf{T}_{j}^{\text{exp}} &:= \mathbf{I} + \Delta t \mathbf{A}_j
                                    + \Delta t \mathbf{D}_j^h \in \mathbb{R}^{n_x \times n_x}, \\
   \label{eqn:tmimp}
    \mathbf{T}_{j}^{\text{imp}} &:= \left( \mathbf{I} - \Delta t \mathbf{D}_j^v
                                    \right)^{-1} \in \mathbb{R}^{n_x \times n_x}
  \end{align}
  for each time instant $t_j$, $j \in \{0, \ldots, n_t - 1\}$. The implicit
  matrices $\mathbf{T}_{j}^{\text{imp}}$, $j \in \{0, \ldots, n_t - 1\}$, are
  block-diagonal again since the inversion of a matrix keeps this structure.
  Finally, a time step of the marine ecosystem model using the TMM is given by
  \begin{align}
    \label{eqn:TMM}
    \vec{y}_{j+1} &= \mathbf{T}_{j}^{\text{imp}}
                     \left( \mathbf{T}_{j}^{\text{exp}} \vec{y}_j
                        + \Delta t \vec{q}_j \left( \vec{y}_j, \vec{u} \right)
                        \right)
                 =: \varphi_j \left( \vec{y}_j, \vec{u} \right),
        \quad j = 0, \ldots, n_t - 1.
  \end{align}
  Due to the grid-point based ocean circulation model, both the explicit and the
  implicit transport matrices are sparse. For the implicit ones, this is given
  by their block-diagonal structure.

  In practical computations, the TMM computed and stored monthly averaged
  matrices and interpolated those linearly for any time instant $t_j$, $j = 0,
  \ldots, n_t - 1$. In this paper, we applied transport matrices computed with
  the MIT ocean model \cite{MAHPH97} used a global configuration with a
  latitudinal and longitudinal resolution of $2.8125^\circ$ and 15 vertical
  layers.

\subsection{Computation of steady annual Cycles}
\label{sec:SteadyAnnualCycles}

  For a marine ecosystem model, the steady annual cycle is a fixed-point of the
  spin-up. Applying the above iteration \eqref{eqn:TMM} over one model year,
  the steady annual cycle (i.e., an annual periodic solution) in a fully
  discrete setting fulfills
  \begin{align}
    \vec{y}_{n_t} &= \vec{y}_0.
  \end{align}
  The steady annual cycle is a fixed-point of the nonlinear mapping
  \begin{align}
    \Phi &:= \varphi_{n_t -1} \circ \ldots \circ \varphi_{0},
  \end{align}
  with $\varphi_j$ defined in \eqref{eqn:TMM}, describing the time integration
  of \eqref{eqn:TMM} over one model year. A classical fixed-point iteration
  takes the form
  \begin{align}
    \label{eqn:Spin-upIteration}
    \vec{y}^{\ell + 1} &= \Phi \left( \vec{y}^{\ell}, \vec{u} \right),
                          \quad \ell = 0, 1, \ldots,
  \end{align}
  using an arbitrary start vector $\vec{y}^{0} \in \mathbb{R}^{n_y n_x}$ and
  model parameters $\vec{u} \in \mathbb{R}^{n_u}$. Interpreting the
  fixed-point iteration as pseudo-time stepping or \emph{spin-up}, the vector
  $\vec{y}^{\ell} \in \mathbb{R}^{n_y n_x}$ contains the tracer
  concentrations at the first time instant of the model year $\ell \in
  \mathbb{N}$.

  We tested the numerical convergence of the spin-up (i.e,. of the iteration
  \eqref{eqn:Spin-upIteration}) with the difference between two consecutive
  iterates determined by
  \begin{align}
    \label{eqn:StoppingCriterion}
    \varepsilon_{\ell} := \left\| \vec{y}^{\ell} - \vec{y}^{\ell - 1} \right\|
  \end{align}
  for iteration (model year) $\ell \in \mathbb{N}_0$. For this purpose, we
  quantified this difference with various norms. Namely, we defined a weighted
  Euclidean norm
  \begin{align}
       \left\| \vec{z} \right\|_{2, w} &:= \left( \sum_{i=1}^{n_{y}} 
          \sum_{k=1}^{n_{x}} w_k z_{ik}^2 \right)^{\frac{1}{2}}
  \end{align}
  with weights $w_k \in \mathbb{R}_{> 0}$ for $k \in \left\{1, \ldots, n_{x}
  \right\}$ and $\vec{z} \in \mathbb{R}^{n_{y} n_{x}}$ indexed as
  $\vec{z} = \left( \left( z_{ik} \right)_{k=1}^{n_{x}}
  \right)_{i=1}^{n_{y}}$. If all weights are equal to 1 (i.e., $w_k = 1$
  for $k = 1, \ldots, n_x$), the norm $\left\| \cdot \right\|_{2, w}$
  corresponds to the Euclidean norm $\left\| \cdot \right\|_2$. We denoted by
  $\left\| \cdot \right\|_{2,V}$ the discretized counterpart of the
  $L^2 (\Omega)^{n_{y}}$ norm using the weights $w_k = \left| V_k \right|$,
  $k = 1, \ldots, n_{x}$, with the box volume $\left| V_k \right|$
  corresponding to the grid point $x_k$. In order to consider not only the
  difference at first time instant but for the whole trajectory over one model
  year, we, moreover, defined the weighted Euclidean norm $\left\| \cdot
  \right\|_{2, w, T}$ by
  \begin{align}
      \left\| \vec{z} \right\|_{2, w, T} &:= \left( \sum_{i=1}^{n_{y}}
        \sum_{j=0}^{n_{t}-1} \Delta t \sum_{k=1}^{n_{x}} w_k z_{jik}^2
        \right)^{\frac{1}{2}}
  \end{align}
  for a weight vector $\vec{w} \in \mathbb{R}^{n_{x}}_{>0}$ and $\vec{z}
  \in \mathbb{R}^{n_{t} n_{y} n_{x}}$ indexed as $\vec{z} = \left( \left(
  \left( z_{jik} \right)_{k=1}^{n_{x}} \right)_{i=1}^{n_{y}}
  \right)_{j=0}^{n_{t}-1}$. Analogous to the weighted Euclidean norm $\left\|
  \cdot \right\|_{2, w}$, we denoted by $\left\| \cdot \right \|_{2, T}$ the
  Euclidean norm and by $\left\| \cdot \right \|_{2,V,T}$ the weighted
  Euclidean norm $\left\| \cdot \right\|_{2, w, T}$ using weights $w_k =
  \left| V_k \right|$ for $k \in \{1, \ldots, n_x\}$. In addition to the norms
  including all grid points of the discretization $\left( x_k
  \right)_{k=1}^{n_x}$, we restricted norms to horizontal layers of the ocean
  discretization, such as to the upper layer describing the ocean surface. We
  identified with $\left\| \cdot \right\| \left.\right|_L$ for $L \subset \{1,
  \ldots, 15\}$ the restriction of norm $\left\| \cdot \right\|$ to the
  layers selected in the set $L$. For example, $\left\| \cdot \right\| \left.
  \right|_L$ with $L := \{1\}$ restricted the norm to the grid points of the
  ocean surface.

\subsection{Temporal Coarsening of Transport Matrices}
\label{sec:TemporalCoarsening}

  Using simple matrix operations, transport matrices computed with a given
  time step can be used to generate matrices corresponding to a bigger time
  step. The procedure has been described in \cite{Kha07}. Following his
  approach, we used the transport matrices $\mathbf{T}_{j}^{\text{exp}}$ and
  $\mathbf{T}_{j}^{\text{imp}}$, $j \in \{0, \ldots, n_t -1 \}$, to generate
  transport matrices
  \begin{align}
    \mathbf{T}_{j, m}^{\text{exp}} &:= \mathbf{I} + m \left(
                            \mathbf{T}_{j}^{\text{exp}} - \mathbf{I} \right), \\
    \mathbf{T}_{j, m}^{\text{imp}} &:= \left( \mathbf{T}_{j}^{\text{imp}}
                                       \right)^m
  \end{align}
  corresponding to a time step with coarsening factor $m \in \mathbb{N}$. 
  Consequently, these matrices represent larger time steps than the ones in the
  underlying ocean circulation model which were used to generate the transport
  matrices $\mathbf{T}_{j}^{\text{exp}}$ and $\mathbf{T}_{j}^{\text{imp}}$. The
  explicit transport matrix $\mathbf{T}_{j, m}^{\text{exp}}$ is the exact
  representation of the larger time step, i.e.,
  \begin{align}
    \mathbf{T}_{j, m}^{\text{exp}}
      &= \mathbf{I} + m \left( \mathbf{T}_{j}^{\text{exp}} - \mathbf{I} \right)
       = \mathbf{I} + m \left( \mathbf{I} + \Delta t \mathbf{A}_j
                                + \Delta t \mathbf{D}_j^h - \mathbf{I} \right)
       = \mathbf{I} + m \Delta t \mathbf{A}_j + m \Delta t \mathbf{D}_j^h.
  \end{align}
  The implicit transport matrix $\mathbf{T}_{j, m}^{\text{imp}}$ is an
  approximation with a loss of accuracy of order $\Delta t^{2}$ since (using
  the binomial theorem)
  \begin{align}
    \mathbf{T}_{j, m}^{\text{imp}}
     &= \left( \mathbf{T}_{j}^{\text{imp}} \right)^m
      = \left( \left( \mathbf{I} - \Delta t \mathbf{D}_j^v \right)^{-1}
        \right)^m
      = \left( \left( \mathbf{I} - \Delta t \mathbf{D}_j^v \right)^m
        \right)^{-1}
     = \left( \sum_{k=0}^m \binom{m}{k} \mathbf{I}^{m-k} \left( -\Delta t
               \mathbf{D}_j^v \right)^k \right)^{-1} \\
    &= \left( \binom{m}{0} \mathbf{I}^m \left( -\Delta t
               \mathbf{D}_j^v \right)^0 + \binom{m}{1} \mathbf{I}^{m-1}
               \left( -\Delta t \mathbf{D}_j^v \right)^1
               + \sum_{k=2}^m \binom{m}{k} \mathbf{I}^{m-k} \left( -\Delta t
               \mathbf{D}_j^v \right)^k \right)^{-1} \\
    &= \left( \mathbf{I} - m \Delta t \mathbf{D}_j^v
              + \sum_{k=2}^m \binom{m}{k} \mathbf{I}^{m-k} \left( -\Delta t
               \mathbf{D}_j^v \right)^k \right)^{-1}
      = \left( \left( \mathbf{I} - m \Delta t \mathbf{D}_j^v \right)
               + \mathcal{O}(\Delta t)^2 \right)^{-1},
  \end{align}
  cf. \cite{PiwSla16a, Kha07}. These transport matrices are still sparse
  \cite{Kha07}.

  The time step of the ocean circulation model used for the computation of the
  transport matrices $\mathbf{T}_{j}^{\text{exp}}$ and
  $\mathbf{T}_{j}^{\text{imp}}$, $j \in \{0, \ldots, n_t -1 \}$, corresponds
  to 3~\si{h} \cite{MAHPH97}. Assuming 360 days per model year,
  the number of time steps per model year is $n_t = 2880$. We, hereinafter,
  denoted this time step with 1 $\Delta t$ using $\Delta t = \tfrac{1}{2880}$. 
  More specifically, time step 2 $\Delta t$ corresponds to the coarsening
  factor 2, i.e., a doubling of the effective time step, with $n_t = 1440$. In
  order to identify the time step used to run the spin-up
  \eqref{eqn:Spin-upIteration}, we denoted by
  \begin{align}
    \vec{y}^{\ell + 1, m} &= \Phi^m \left( \vec{y}^{\ell}, \vec{u} \right),
    \quad \ell = 0, 1, \ldots,
  \end{align}
  the time-integration of one model year using the time step $m \Delta t$ with
  the factor $m \in \mathbb{N}$ for $\vec{y}^{\ell} \in \mathbb{R}^{n_y n_x}$
  and $\vec{u} \in \mathbb{R}^{n_u}$. More importantly, the transport matrices
  $\mathbf{T}_{j, m}^{\text{exp}}$ and $\mathbf{T}_{j, m}^{\text{imp}}$,
  $j \in \{0, \ldots, n_{t} - 1\}$, enter \eqref{eqn:TMM} for the application
  of the time step $m \Delta t$ instead of the transport matrices
  $\mathbf{T}_{j}^{\text{exp}}$ and $\mathbf{T}_{j}^{\text{imp}}$.
  
  Any choice $m \in \mathbb{N}$ is possible in the above computations. For every
  value of $m$ to be used, twelve pairs of explicit and implicit transport
  matrices, however, have to be supplied. Since we wanted to check a rather wide
  variability of the time steps, we have chosen the possible coarsening factors
  \begin{align}
    \label{def:M}
    m \in M &:= \{1, 2, 4, 8, 16,  32, 64\}
  \end{align}
  in this study. The upper limit of $m = 64$ was motivated by our observation
  that the spin-up for most of the models did not converge anymore for larger
  time steps.

\subsection{Negative Tracer Concentrations}
\label{sec:NegativeTracerConcentrations}

  A time step that is too large can cause negative tracer concentrations. Given
  a non-negative tracer distribution $\vec{y}_j$ in all grid points at a time
  instant $t_j$, $j \in \{0, \ldots, n_t - 1\}$, the multiplication with the
  explicit transport matrix $\mathbf{T}_{j}^{\text{exp}}$ in \eqref{eqn:TMM}
  will always result in a non-negative distribution. In contrast, the source
  term $\vec{q}_j \left( \vec{y}_j, \vec{u} \right)$ may give negative values
  for some tracers at some points even though the input variable $\vec{y}_j$ is
  non-negative. On the other hand, negative tracer concentrations do not make sense. 
  Moreover and without going into mathematical details, it can be shown that 
  at least in the semi-discrete, time-continuous setting, a non-negative initial value always leads to a non-negative tracer distribution for all time, at least for the biogeochemical models used here. Thus, negative values may be obtained only due to a time step that is too big. Following this idea, one might consider to guarantee non-negativity throughout the simulation, in order to keep the solution consistent with biogeochemistry and mathematics.
   In many models, non-negativity is enforced by a simple setting of
  negative values to zero before the evaluation of the source term. Such a
  setting obviously increases the total mass in the ecosystem. Hence, a frequent
  occurrence of big negative values and their correction to zero will result in
  a changed steady solution obtained in the spin-up. Despite such a
  correction before the source term step, the sum in the bracket in
  \eqref{eqn:TMM} might be negative at some points, depending on the used time
  step. Thus, after applying the implicit matrix, the result of the time step
  $\vec{y}_{j+1}$ might contain negative values as well.

  Small negative values in some points do not necessarily change the convergence
  of a spin-up to a steady annual cycle. Therefore, a main criterion for
  algorithms that use bigger time steps (as the ones presented here) is if the
  spin-up converges, and if it converges to the same solution as for the
  standard setting. We also investigated an algorithm that enforces non-negative
  tracer concentrations and how this effects the solution and the cost saving.

%% file: Latexfiles/StepSizeControl_arxiv.tex
Methods for automatic step size control automatically adapt the time step during
the calculation of a transient computation. They may, therefore, also be used
for the steady annual cycle computation via a spin-up. A step size
control method estimates the local discretization error by computing two
approximations, either with different time steps (in the method used here) or
by using two different time integration methods. The approach used here is based
on the Richardson extrapolation \cite{RicGla11, RicGau27}. For the estimate of
the local discretization error, the step size control computes two
approximations with two different time step sizes in every step. Depending on
the error estimate and a desired accuracy, the step size control accepts or
rejects the approximation calculated with the smaller time step (the
approximation computed with the larger time step always serves the error
estimation only). Then, it adapts the step size to a value that, using the
estimation of the error, would result in the desired accuracy. Thus, in this
step, an increase or a decrease of the step size is possible. Using the adapted
step size, the method, afterwards, either starts the calculation of the next
step or reruns the calculation of the current step. In summary, the step size
control finally uses always the largest time step that keeps the error below the
given tolerance.

Step size control methods are based on the existence of an asymptotic expansion
of the discretization error \cite[Sect. 7.2.3]{StoBul02}. To obtain this
result, the unknown solution is assumed to be three times continuously
differentiable in time. The question whether this regularity is satisfied for
the models investigated in this work is not discussed here. We just note that
all models have continuous right-hand sides (i.e., source-minus-sink terms).
Furthermore, it is quite usual to apply a step size control even though some
assumptions of the underlying mathematical theory may not be given or shown for
an actual application. Since we compared our results obtained with the step size
control to those with constant time steps, an assessment of the method is
possible.

\input{Latexfiles/Algorithm.tex}
Algorithm \ref{alg:StepSizeControl} depicts the step size control for the
calculation of a steady annual cycle. It is based on
\cite[Algorithm 5.2]{DeuBor13}. Since we are aiming at a steady annual cycle
obtained in a spin-up, we do not check for the local discretization error after
every single time step, but only after an adjustable number of model years. This
number ($n_{s} \in \mathbb{N}$) is one of the input parameters of the algorithm.
The loop over the $n_{s}$ years is realized in lines \ref{alg:StartModelYears}
to \ref{alg:EndModelYears}. Another important feature is the set of possible or
desired multiplication factors $m$. Here, the  value  $m = \max M = 64$ is inadmissible
because the step size control estimates the local error using two approximations
calculated with the selected time step $m \Delta t$ and a larger time step.
Thus, the set $\hat M:=\{1,2,4,8,16,32\}$ defines the admissible time steps of
Algorithm \ref{alg:StepSizeControl}.

\input{Latexfiles/AlgorithmStepSizeControlAvoiding.tex}

We designed the step size control algorithm having various options. The user may
select, firstly, the initial time step $m_{\text{init}} \Delta t$ of the step
size control with $m_{\text{init}} \in \hat M $. The second option is
the number $n_{s} \in \mathbb{N}$ of model years after which the error is
estimated. The step size control computes the two approximations with time
steps $m \Delta t$ and $\tilde{m} \Delta t$ for the model year $\ell + n_s$,
$\ell \in \mathbb{N}_0$, and $m, \tilde{m} \in M$ with $m < \tilde{m}$ and
estimates the local discretization error $\left\| \mathbf{y}^{\ell + n_s, m} -
\mathbf{y}^{\ell + n_s, \tilde{m}} \right\|$. The choice of the next time step
depends on this error (line \ref{alg:ErrorEstimation} or
\ref{alg:ErrorEstimationElse} and following). Thirdly, it is possible to choose
the norm that may have an influence on the error estimation. Another setting of
the step size control, fourthly, is the tolerance $\tau_0 \in \mathbb{R}_{>0}$
which controls the acceptance of the approximation. By default, the step size
control applies the coarsening factor $m_{\text{init}} = 1$ for the initial
time step $m_{\text{init}} \Delta t$, $n_s = 1$, the volume-weighted Euclidean
norm $\left\| \cdot \right\|_{2, V}$ and the tolerance $\tau_0 = 1$.

A variant of Algorithm \ref{alg:StepSizeControl} includes an additional
avoidance of negative concentrations (see Algorithm
\ref{alg:StepSizeControlAvoiding}). If enabled, the step size control accepts an
approximation -- as additional criterion to the local discretization error -- if
and only if the concentration of this approximation is non-negative at all
grid points $x_k, k \in\{1, \ldots, n_x\}$, or if the approximation was
calculated with the smallest possible time step. Otherwise, the step size
control reruns this step with a decreased time step.

%% file: Latexfiles/Algorithm.tex
\renewcommand{\algorithmicrequire}{\textbf{Input:}}
\renewcommand{\algorithmicensure}{\textbf{Output:}}
\renewcommand{\algorithmiccomment}[1]{\unskip\hfill\makebox[.55\textwidth][l]{\texttt{// #1}}\par}

\begin{algorithm}[!tb]
  \caption{Step size control.}
  \label{alg:StepSizeControl}
  \begin{algorithmic}[1]
    \REQUIRE{ Initial concentration: $\mathbf{y}^{0} \in \mathbb{R}^{n_y n_x}$, \\
      \hspace{1.2\algorithmicindent} Parameter vector: $\mathbf{u} \in \mathbb{R}^{n_u}$, \\
      \hspace{1.2\algorithmicindent} Number of model years: $T \in \mathbb{N}$, \\
      \hspace{1.2\algorithmicindent} Coarsening factor of the initial time step: $m_{\text{init}} \in \hat M:= M\setminus \max M$, \\
      \hspace{1.2\algorithmicindent} Number of model years for the error estimation: $n_s \in \mathbb{N}$, \\
      \hspace{1.2\algorithmicindent} Tolerance: $\tau_0 \in \mathbb{R}_{>0}$}
    \ENSURE{$\mathbf{y}^{T} \in \mathbb{R}^{n_y n_x}$} \text{(superscript referring to $\ell=T$ here, not meaning transposed vector)}
    \STATE $\ell = 0$ \label{alg:FirstLine}
    \STATE $\Delta t = \frac{1}{2880}$
    \STATE $m_{\text{min}} = \min M$
    \STATE $m_{\text{max}} = \max \left\{ \bar{m} \in M : \exists p \in M : \bar{m} < p \right\}$
    \STATE $m = m_{\text{init}}$
    \WHILE[Spin-up over (at least) T model years]{$\ell \leq T$}
      \STATE $\tilde{m} = \max \left\{ \bar{m} \in M : \bar{m} \leq 2 \cdot m \right\}$
      \STATE $\mathbf{y}^{\ell, m} = \mathbf{y}^{\ell}$
      \STATE $\mathbf{y}^{\ell, \tilde{m}} = \mathbf{y}^{\ell}$
      \FOR[Compute approximations to estimate the error]{$i = 1$ to $n_s$} \label{alg:StartModelYears}
        \STATE $\mathbf{y}^{\ell + i, m} = \Phi^{m} \left( \mathbf{y}^{\ell + i - 1, m}, \mathbf{u} \right)$
        \STATE $\mathbf{y}^{\ell + i, \tilde{m}} = \Phi^{\tilde{m}} \left( \mathbf{y}^{\ell + i -1, \tilde{m}}, \mathbf{u} \right)$
      \ENDFOR \label{alg:EndModelYears}
      \STATE $\epsilon = 2 \cdot \left\| \mathbf{y}^{\ell + n_s, m} - \mathbf{y}^{\ell + n_s, \tilde{m}} \right\|$ \label{alg:ErrorNorm}
      \IF[Accept approximation and increase time step m]{$\frac{\epsilon}{m \Delta t} \leq \tau_0$ \OR $m = m_{\text{min}}$} \label{alg:ErrorEstimation}
        \STATE $\mathbf{y}^{\ell + n_s} = \mathbf{y}^{\ell + n_s, m}$ \label{alg:AcceptApproximation}
        \STATE $m_{\text{opt}} = \max \left\{ \bar{m} \in M : \bar{m} \leq \left
              \lfloor \frac{\tau_0}{\epsilon} \cdot \left( 
              m \Delta t \right)^{2} \right\rfloor
              \right\}$
        \STATE $m_{\text{scale}} = \max \left\{ \bar{m} \in M : \bar{m} \leq
             2 \cdot m \right\}$
        \STATE $m = \max \left\{ m_{\text{min}}, \min \left\{ m_{\text{max}},
              m_{\text{scale}}, m_{\text{opt}} \right\} \right\}$
        \STATE $\ell = \ell + n_s$
      \ELSE[Reject approximation and decrease time step m] \label{alg:ErrorEstimationElse}
        \STATE $m = \max \left\{ \bar{m} \in M : \bar{m} \leq \left\lfloor \frac{m}{2} \right\rfloor \vee \bar{m} = m_{\text{min}} \right\}$
      \ENDIF
    \ENDWHILE \label{alg:LastLine}
  \end{algorithmic}
\end{algorithm}

%% file: Latexfiles/AlgorithmStepSizeControlAvoiding.tex
\renewcommand{\algorithmiccomment}[1]{\texttt{// #1}}
\renewcommand{\algorithmiccomment}[1]{\unskip\hfill\makebox[.95\textwidth][l]{\texttt{// #1}}\par}

\begin{algorithm}[tb]
  \caption{Step size control avoiding negative concentrations. Shown are only  differences to Algorithm
  \ref{alg:StepSizeControl}.}
  \label{alg:StepSizeControlAvoiding}
  \begin{algorithmic}[1]
    \REQUIRE{Initial concentration: $\mathbf{y}^{0} \in \mathbb{R}^{n_y n_x}$, \\
         \hspace{1.2\algorithmicindent} Parameter vector: $\mathbf{u} \in \mathbb{R}^{n_u}$, \\
         \hspace{1.2\algorithmicindent} Number of model years: $T \in \mathbb{N}$, \\
         \hspace{1.2\algorithmicindent} Coarsening factor of the initial time step: $m_{\text{init}} \in M$, \\
         \hspace{1.2\algorithmicindent} Number of model years for the error estimation: $n_s \in \mathbb{N}$, \\
         \hspace{1.2\algorithmicindent} Tolerance: $\tau_0 \in \mathbb{R}_{>0}$}
    \ENSURE{$\mathbf{y}^{T} \in \mathbb{R}^{n_y n_x}$}
    
    \COMMENT{Lines \ref{alg:FirstLine} to \ref{alg:ErrorNorm} identical as in Algorithm \ref{alg:StepSizeControl}}
      \setcounter{ALC@line}{14}
      \IF{$\frac{\epsilon}{m \Delta t} \leq \tau_0$ \AND $\big( \forall j \in \{1, \ldots, n_y\} \forall i \in \{1, \ldots, n_x\}: \mathbf{y}^{\ell + n_s, m}_{ji} \geq 0 \big)$ \OR $m = m_{\text{min}}$}
      
        \STATE \COMMENT{Lines \ref{alg:AcceptApproximation} to \ref{alg:LastLine} identical as in Algorithm
        \ref{alg:StepSizeControl}}
        \setcounter{ALC@line}{22}
      \ENDIF
  \end{algorithmic}
\end{algorithm}

%% file: Latexfiles/DecreasingTimesteps.tex
The step size control algorithms described above require an increased
computational effort to automatically adjust the time step. They both try
to find the optimal step size in the sense that it should be as big as possible
but still keeps the local discretization error below the given tolerance.
Therefore, the algorithms are automatically able to both increase and decrease
the time step during the spin-up. This optimality comes with the additional
effort of computing always two approximations to estimate this error.

\input{Latexfiles/AlgorithmDecreasingTimesteps.tex}
The decreasing time steps algorithm presented in this section, in contrast,
exclusively decreases the time step during the spin-up. The motivation is the
same, namely to reduce the computational costs. The algorithm assumes that in
its beginning bigger time steps are sufficient. The procedure is shown in
Algorithm \ref{alg:DecreasingTimesteps}. Starting from an initial time step
$m_{\text{init}} \Delta t$ with a coarsening factor $m_{\text{init}} \in M$, the
algorithm uses the same time step until the norm $\left\| \mathbf{y}^{\ell} -
\mathbf{y}^{\ell + n_s} \right\|$, $\ell \in \mathbb{N}_0$, fell below a given
tolerance $\varepsilon \in \mathbb{R}_{>0}$. In other words, the spin-up
calculation with this time step already reached almost a steady annual cycle. If
the norm fell below the tolerance, the algorithm decreases the time step. The
assessment whether a significant reduction was still achieved with the current
time step takes place after a fixed number of $n_s \in \mathbb{N}$ model years.
By default, the decreasing time steps algorithm uses the initial time step
$m_{\text{init}} \Delta t$ with the coarsening factor $m_{\text{init}} = 64$,
$n_s = 50$ and the tolerance $\varepsilon = 0.001$.

%% file: Latexfiles/AlgorithmDecreasingTimesteps.tex
\renewcommand{\algorithmiccomment}[1]{\unskip\hfill\makebox[.6\textwidth][l]{\texttt{// #1}}\par}

\begin{algorithm}[tb]
  \caption{Decreasing time steps.}
  \label{alg:DecreasingTimesteps}
  \begin{algorithmic}[1]
    \REQUIRE{Initial concentration: $\mathbf{y}^{0} \in \mathbb{R}^{n_y n_x}$, \\
         \hspace{1.2\algorithmicindent} Parameter vector: $\mathbf{u} \in \mathbb{R}^{n_u}$, \\
         \hspace{1.2\algorithmicindent} Number of model years: $T \in \mathbb{N}$, \\
         \hspace{1.2\algorithmicindent} Coarsening factor of the initial time step: $m_{\text{init}} \in M$, \\
         \hspace{1.2\algorithmicindent} Number of model years for the error estimation: $n_s \in \mathbb{N}$, \\
         \hspace{1.2\algorithmicindent} Tolerance: $\varepsilon \in \mathbb{R}_{>0}$}
    \ENSURE{$\mathbf{y}^{T} \in \mathbb{R}^{n_y n_x}$}
    \STATE $\ell = 0$
    \STATE $m_{\text{min}} = \min M$
    \STATE $m = m_{\text{init}}$
    \WHILE[Spin-up over (at least) T model years]{$\ell \leq T$}
      \FOR[Spin-up over $n_s$ model years]{$i = 1$ to $n_s$}
        \STATE $\mathbf{y}^{\ell + i} = \Phi^{m} \left( \mathbf{y}^{\ell + i - 1}, \mathbf{u} \right)$
      \ENDFOR
      \IF[Check reduction and decrease time step m]{$\left\| \mathbf{y}^{\ell} - \mathbf{y}^{\ell + n_s} \right\| < \varepsilon$ \AND $m_{\text{min}} < m$}
        \STATE $m = \max \left\{ \bar{m} \in M : \bar{m} < m \right\}$
      \ENDIF
      \STATE $\ell = \ell + n_s$
    \ENDWHILE
  \end{algorithmic}
\end{algorithm}

%% file: Latexfiles/Results_arxiv.tex
The automatic adjustment of the time step used for the spin-up computation
influenced both the computational effort and the accuracy of the approximation
of the steady annual cycle. In this section, we present the numerical
results using the adaptive step size control (see Sect.
\ref{sec:StepSizeControl}) and the decreasing time steps algorithm (Sect.
\ref{sec:DecreasingTimesteps}) to shorten the runtime of the computation of the
steady annual cycle. We assessed the accuracy and cost saving of the calculated
approximations.

\subsection{Experimental Setup}
\label{sec:ExperimentalSetup}

  For each part of the spin-up calculation (i.e., $n_s \in \mathbb{N}$ model
  years), we used the marine ecosystem toolkit for optimization
  and simulation in 3D, Metos3D, \cite{PiwSla16, PiwSla16a}. Overall, we
  computed all spin-ups over 10000 model years. We started the spin-up
  always with a constant global mean tracer concentration of
  $2.17$~\si{mmol\, P\, m^{-3}} for $\textrm{PO}_{4}$ and, if present,
  $0.0001$~\si{mmol\, P\, m^{-3}} for the tracers \textrm{DOP}, \textrm{P},
  \textrm{Z} and \textrm{D}.

  Using different parameter vectors, we calculated the steady annual cycles for
  the various biogeochemical models. We used, on the one hand, the reference
  parameter vectors listed in Table \ref{table:ParameterValues-Modelhierarchy}
  and, on the other hand, 100 parameter vectors generated by a Latin
  hypercube sample within the bounds defined in Table
  \ref{table:ParameterValues-Modelhierarchy} for each biogeochemical model, cf.
  \cite{McBeCo79}. We created these parameter vectors by the
  \texttt{lhs} routine of \cite{PyDOE17}.

  For the assessment of the calculated approximations of the steady annual
  cycle, we compared them with reference solutions. These were chosen as the
  result obtained by a spin-up using Metos3D with constant time step
  $m\,\Delta t, m\in M$, also over 10000 model years. These solutions are
  denoted by $\mathbf{y}^{10000, m}$ and described in more detail in
  \cite{PfeSla21a}. In most cases, we considered the case $m=1$ only. The
  reference solutions were also used to measure, in particular, the accuracy of
  an approximation $\mathbf{x} \in \mathbb{R}^{n_y n_x}$ obtained by one of the
  algorithms (i.e., the step size control or the decreasing time steps
  algorithm) by the relative difference
  \begin{align}
    \label{eqn:relativeError}
    \frac{\left\| \mathbf{x} - \mathbf{y}^{10000, 1} \right\|_2}
         {\left\| \mathbf{y}^{10000, 1} \right\|_2}.
  \end{align}
  We call this quantity \eqref{eqn:relativeError} the \emph{(relative) error}
  of the respective result $\mathbf{x}$. Furthermore, we quantified the
  saving of the computational costs by
  \begin{align}
    \label{eqn:CostSavings}
    \frac{c_{\text{ref}} - c_{\text{algo}}}{c_{\text{ref}}},
  \end{align}
  where $c_{\text{ref}} \in \mathbb{N}$ denotes the number of model evaluations
  used to compute the reference solution (i.e., $c_{\text{ref}} = 10000 \cdot
  2880$, according to the $n_t = 2880$ time steps for each of the 10000
  model years) and $c_{\text{algo}} \in \mathbb{N}$ denotes the number of model
  evaluations of the respective algorithm.

\subsection{Step Size Control Algorithms}
\label{sec:Results-StepSizeControl}

  In this section, we present the results using the step size control to
  compute a steady annual cycle (Algorithms \ref{alg:StepSizeControl} and
  \ref{alg:StepSizeControlAvoiding}). The step size control applies different
  time steps to adapt the step size using an estimation of the local
  discretization error. Starting with a default setting for Algorithm
  \ref{alg:StepSizeControl} (Sect. \ref{sec:Results-StepsizeControlDefault}),
  we also analyzed the behavior using the avoidance of negative tracer
  concentrations using Algorithm \ref{alg:StepSizeControlAvoiding} (Sect.
  \ref{sec:Results-StepSizeControl-AvoidingNegativeConcentrations}) and,
  briefly, other configuration settings (Sect.
  \ref{sec:Results-StepSizeControl-StepSizeControlSettings}) defined in Sect.
  \ref{sec:StepSizeControl}.

  \subsubsection{Algorithm \ref{alg:StepSizeControl} with default Step Size Control Setting}
  \label{sec:Results-StepsizeControlDefault}

    \begin{figure}[tbp]
      \centering
      \includegraphics{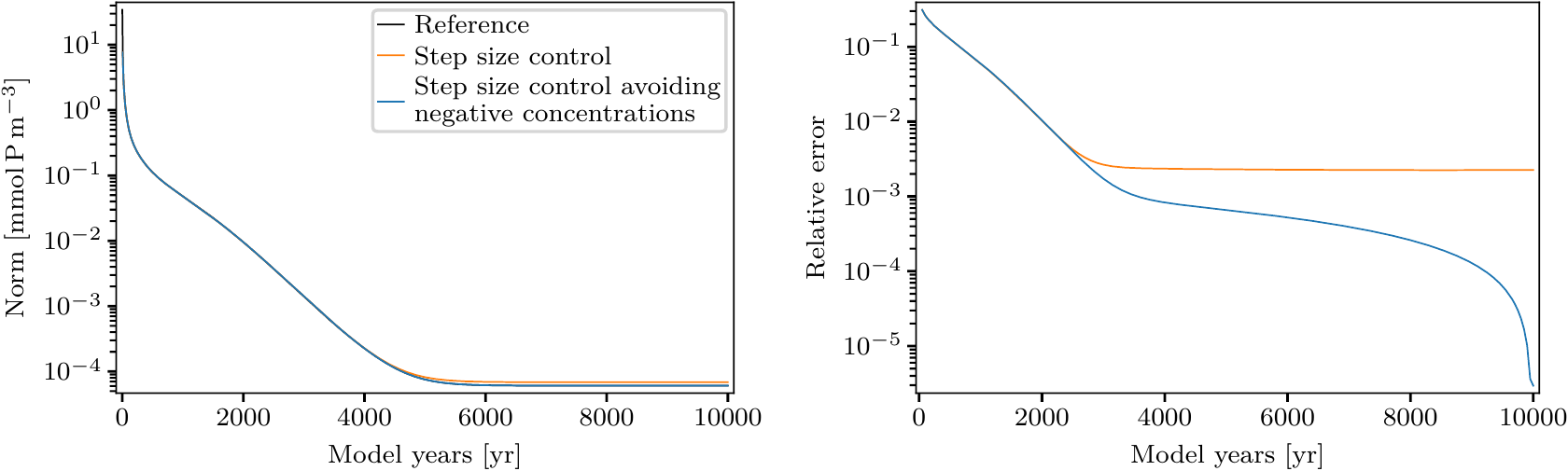}
      \caption{Results of spin-up computations for the reference parameter
               vector (see Table \ref{table:ParameterValues-Modelhierarchy})
               and the N model using the step size control with default setting (Algorithm \ref{alg:StepSizeControl}) and with avoidance of
               negative concentrations (Algorithm
               \ref{alg:StepSizeControlAvoiding}). Shown are the convergence of
               the spin-up (left), i.e., the norms of the difference
               \eqref{eqn:StoppingCriterion} between consecutive iterations, and
               the relative error \eqref{eqn:relativeError} (right).
               On the left, the curves for the reference run and the one with
               step size control avoiding negative concentrations overlap.}
      \label{fig:StepSizeControl_Convergence}
    \end{figure}

    For the reference parameter vectors of Table
    \ref{table:ParameterValues-Modelhierarchy}, the step size control algorithm
    with default setting computed approximations of the steady annual cycle that
    were almost identical to the ones obtained with a constant step size. For
    the N, N-DOP and MITgcm-PO4-DOP models, the step size control utilized the
    largest possible time step by increasing the time step in the first steps to
    32\,$\Delta t$. Accordingly, the algorithm lowered the computational costs
    by 95\%. In contrast, the step size control used always time step
    1\,$\Delta t$ (i.e., no increase of the time step occurred) for the other
    three biogeochemical models NP-DOP, NPZ-DOP and NPZD-DOP. Therefore, the
    computational effort for these three models was greater than for the
    solution $\mathbf{y}^{10000, 1}$. The value computed with formula
    \eqref{eqn:CostSavings} was -0.5. Obviously, the obtained solution was the
    same as the one using constantly 1\,$\Delta t$. As an example, Fig.
    \ref{fig:StepSizeControl_Convergence} indicates the similar convergence behavior
    towards a steady annual cycle using the step size control and the constant
    step size 1\,$\Delta t$, both for the N model. The convergence behavior
    using the step size control concurred with the one of the spin-up with
    constant time step 32\,$\Delta t$, i.e. $\mathbf{y}^{10000, 32}$. In this
    case, the step size control doubled the time step five times up to the
    maximal time step 32\,$\Delta t$ in the first five model years. Afterwards,
    the time step remained unchanged for the entire spin-up, cf.
    \cite[Figs. 1 and 4]{PfeSla21a}. Consequently, the accuracy of this
    approximation resembled that of $\mathbf{y}^{10000, 32}$ (Fig.
    \ref{fig:StepSizeControl_Convergence}, right). The step size control
    resulted in a cost saving of 95\% compared to the solution
    $\mathbf{y}^{10000, 1}$. Indeed, the cost saving was even 97\% using
    directly the spin-up with time step 32\,$\Delta t$ because the step size
    control necessitated two approximations calculated with different time steps
    to estimate the local error for every model year. Looking at the reduction,
    it would be possible to terminate the step size control even after a much
    lower number of model years to lower further the computational effort (Fig.
    \ref{fig:StepSizeControl_Convergence}). For the N-DOP and MITgcm-PO4-DOP
    model, the results were similar.

    \begin{figure}[!tb]
      \centering
      \includegraphics{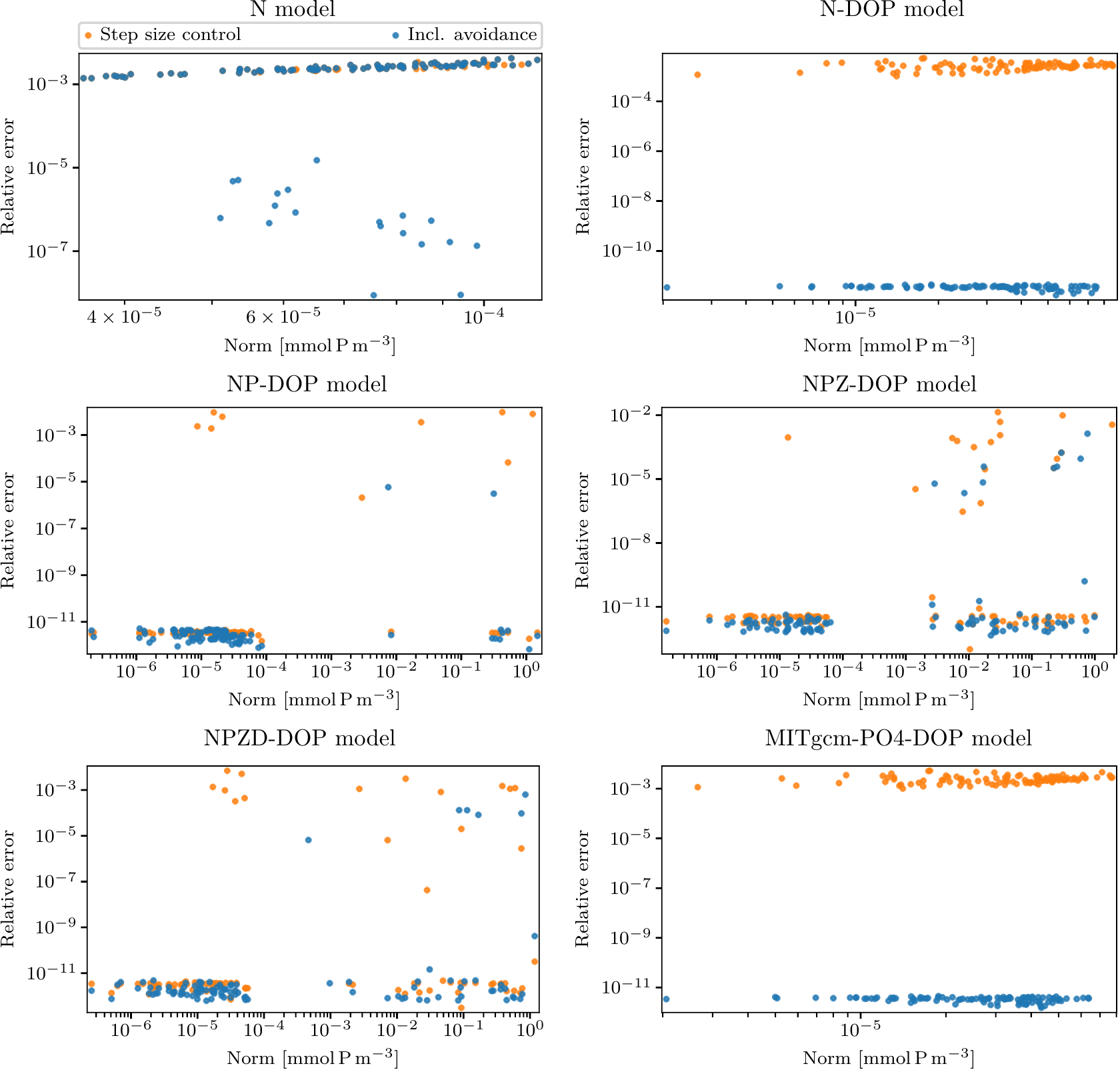}
      \caption{Norm of difference \eqref{eqn:StoppingCriterion} and relative error
               \eqref{eqn:relativeError} for $\ell = 10000$ using the step size
               control with the default setting and with avoidance of negative
               concentrations. Shown are the results for the different
               biogeochemical models and all parameter vectors of the Latin
               hypercube sample. Some
               points of the results using the avoidance of negative
               concentrations obscured the points of the step size control
               results.}
      \label{fig:StepSizeControl_Scatter}
    \end{figure}

    \begin{figure}[!tb]
      \centering
      \includegraphics{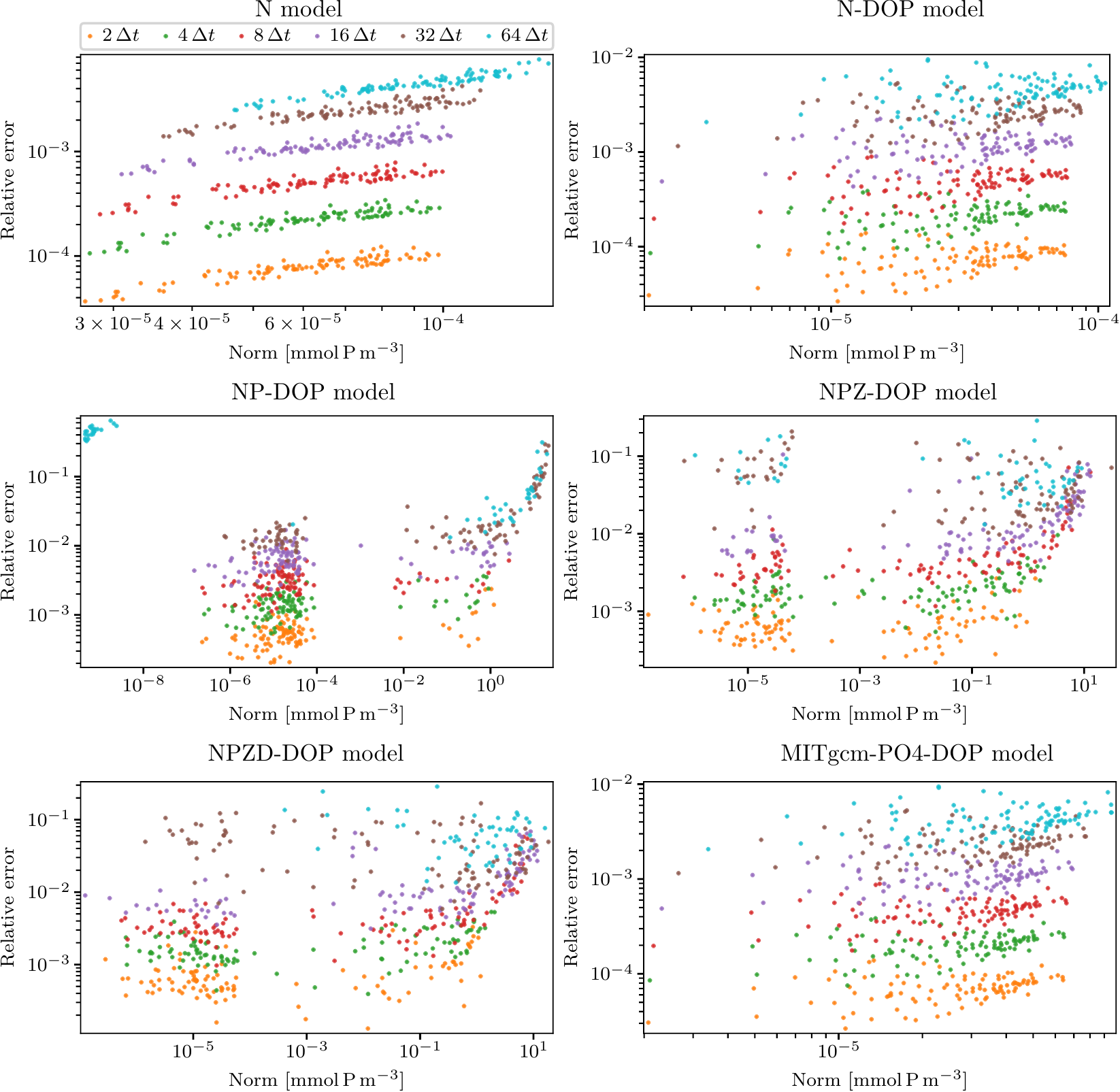}
      \caption{Norm of difference \eqref{eqn:StoppingCriterion} and relative
               error \eqref{eqn:relativeError} for $\ell = 10000$ using
               different time steps for the spin-up. Shown are the results using
               a constant time step over the whole spin-up for the different
               biogeochemical models and all parameter vectors of the Latin
               hypercube sample.}
      \label{fig:Timesteps_Convergence}
    \end{figure}

    \begin{figure}[!tb]
      \centering
      \includegraphics{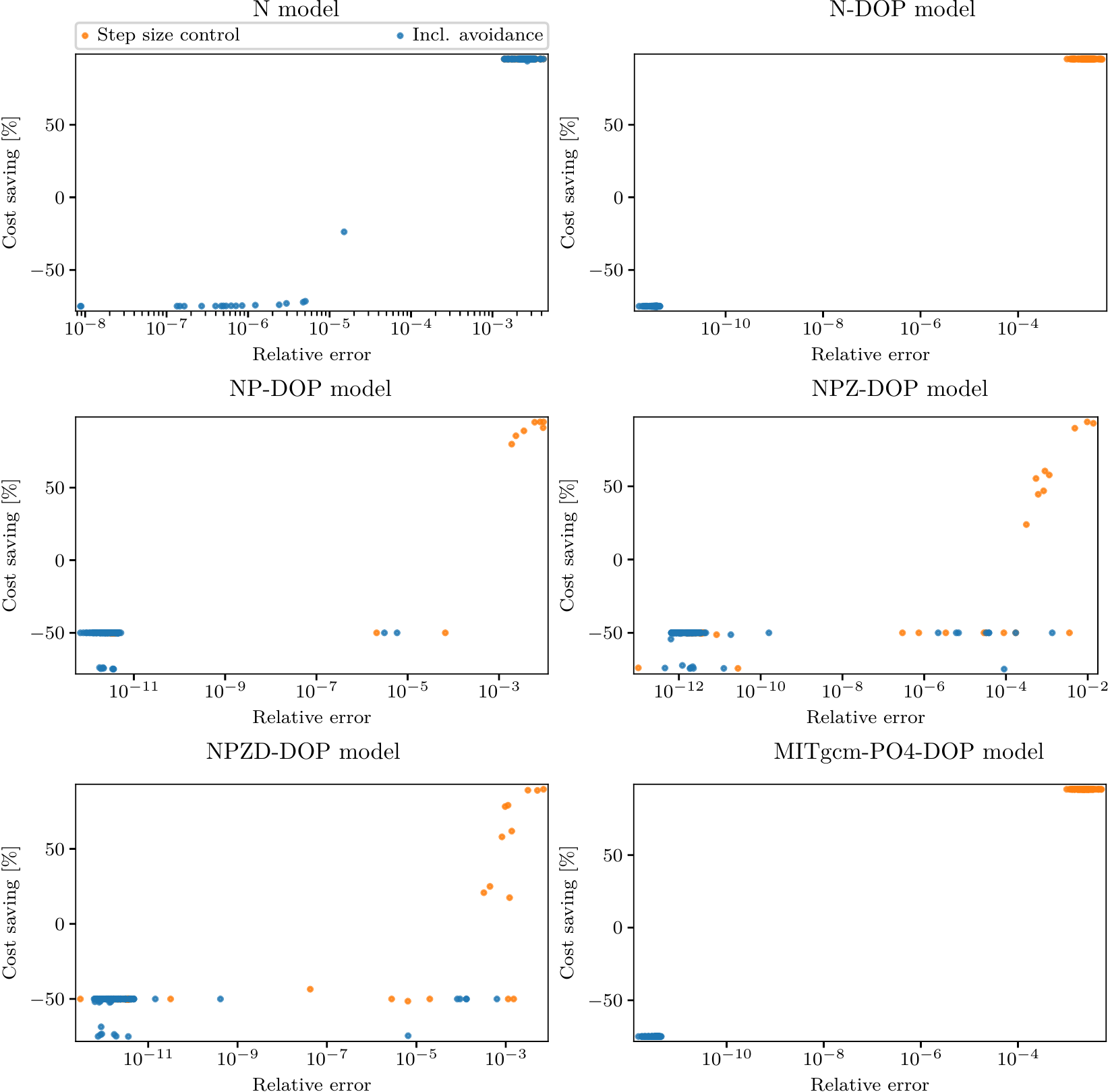}
      \caption{Saving of computational costs \eqref{eqn:CostSavings}
               using the step size control with the default setting and the
               avoidance of negative concentrations. Shown are the results for
               the different biogeochemical models and all parameter vectors
               of the Latin hypercube sample. Some points of the results using the avoidance of
               negative concentrations obscured the points of the step size
               control results. For the N model, all points of the results using
               the step size control (orange points) are present in the upper
               right corner.}
      \label{fig:StepSizeControl_CostSaving}
    \end{figure}

    Also for the 100 parameter vectors of the Latin hypercube sample, Algorithm
    \ref{alg:StepSizeControl} yielded solutions that showed the same accuracy as
    solutions obtained with a fixed step size. For the N, N-DOP and
    MITgcm-PO4-DOP models, the results were consistent with those for the
    reference parameter vectors above, i.e., the accuracy of the approximations
    coincided with that for the spin-up calculated with constant time step
    32\,$\Delta t$ (Figs. \ref{fig:StepSizeControl_Scatter} and
    \ref{fig:Timesteps_Convergence}). The algorithm increased the time step up
    to the maximum of 32\,$\Delta t$ and, afterwards, used only this time step.
    For a few parameter vectors, the step size control, however, did not
    increase the time step directly, but computed several model years before
    increasing the time step. For these three models, the step size control
    resulted in a tremendous cost saving (Fig.
    \ref{fig:StepSizeControl_CostSaving}). For the NP-DOP, NPZ-DOP and NPZD-DOP
    models and more than 85 parameter vectors, the algorithm did not increase
    the time step. Thus, the approximations corresponded to the reference
    solution $\mathbf{y}^{10000,1}$ (Fig. \ref{fig:StepSizeControl_Scatter}) in
    these cases. For the remaining parameter vectors, at least one increase and,
    possibly, some decreases of the time step took place. This resulted in a
    larger relative error but still a reasonable accuracy of the calculated
    approximation. Except for a few parameter vectors where the step size
    control increased the time step, the step size control caused a larger
    computational effort than using 1\,$\Delta t$ constantly because the
    algorithm needs two approximations to estimate the local discretization
    error (Fig. \ref{fig:StepSizeControl_CostSaving}).

  \subsubsection{Algorithm \ref{alg:StepSizeControlAvoiding}: Avoiding negative Concentrations}
  \label{sec:Results-StepSizeControl-AvoidingNegativeConcentrations}
  
    Strictly avoiding negative concentrations prevented a usage of larger time
    steps in most of the considered cases when calculating a steady annual cycle
    with the step size control. Except for the N model, Algorithm
    \ref{alg:StepSizeControlAvoiding} used time step 1\,$\Delta t$ for the
    entire spin-up with the reference parameter vectors of Table
    \ref{table:ParameterValues-Modelhierarchy} because negative concentrations
    appeared for at least one tracer in all approximations calculated with
    larger time steps. For the N model, the algorithm increased the time step up
    the maximum at the beginning of the simulation, but accepted only time step
    1\,$\Delta t$ after 120 model years because negative concentrations arose.
    Accordingly, the approximation resembled the reference solution while the
    computational effort was huge (cost increase of 73\%, Fig.
    \ref{fig:StepSizeControl_Convergence}). For the NP-DOP, NPZ-DOP and NPZD-DOP
    models, the computations using the step size control with and without
    avoidance of negative concentrations were identical. For the N-DOP and
    MITgcm-PO4-DOP models, in contrast, the computational effort was quite big
    (an increase of about 75\%). Here, a bigger time step would be possible
    without considering negative values due to the error estimation. With our implementation of the step size control avoiding negative concentrations,
    three instead of two approximations, hence, were computed for each model
    year because the step size control accepted in every step the approximation
    calculated with the minimal time step 1\,$\Delta t$ (independent of negative
    concentrations) and increased based on the error estimation the time step
    for the next step. In the next step, the algorithm discarded the approximation
    calculated with the larger time step due to negative concentrations, and
    reran the approximation with the minimal time step 1\,$\Delta t$ again. With
    an adjustment of our implementation (firstly, the calculation of an
    approximation using the small time step to check it for negative
    concentrations and, secondly, the computation of the second approximation
    using the larger time step to estimate the local error), the computational
    effort could slightly be reduced to an increase to about 50\%.

    For the Latin hypercube sample, the same qualitative results were obtained.
    For the N model, the results mostly coincided with the results of the step
    size control using default settings (Fig. \ref{fig:StepSizeControl_Scatter}).
    Only for a few parameter vectors, the step size control reduced the time
    step. This was again a result of negative concentrations during the
    simulation. As a consequence, the accuracy improved but at the expense of
    the cost saving (Fig. \ref{fig:StepSizeControl_CostSaving}). For the other
    biogeochemical models, the accuracy of the approximations in Fig.
    \ref{fig:StepSizeControl_Scatter} indicates that the Algorithm
    \ref{alg:StepSizeControlAvoiding} computed the reference solution except
    for some outliers because the step size control did not increased the time
    step during the simulation. Thus, the computational effort was larger than
    for the spin-up computation of the reference solution with 1\,$\Delta t$
    (Fig. \ref{fig:StepSizeControl_CostSaving}).

  \subsubsection{Step size control settings}
  \label{sec:Results-StepSizeControl-StepSizeControlSettings}

    We investigated the influence of different settings for the step size
    control. For that purpose, we used only the reference parameter vectors
    listed in Table \ref{table:ParameterValues-Modelhierarchy}.

    \begin{table}[tb]
      \caption{Relative error \eqref{eqn:relativeError} for $\ell = 10000$
               obtained by the step size control algorithm (Algorithm
               \ref{alg:StepSizeControl}) using different values of $n_s$ and the reference parameter vectors taken from Table \ref{table:ParameterValues-Modelhierarchy}.}
      \label{table:StepSizeControlStepYear}
      \centering
      \begin{tabular}{c c c c c c c}
        \hline
        $n_s$ & N & N-DOP & NP-DOP & NPZ-DOP & NPZD-DOP & MITgcm-PO4-DOP \\
        \\hline
         1 & 2.258e-03 & 2.775e-03 & 3.439e-12 & 1.744e-12 & 1.845e-12 & 2.462e-03 \\
         2 & 2.258e-03 & 2.775e-03 & 7.978e-13 & 7.399e-13 & 5.247e-13 & 2.462e-03 \\
         5 & 2.258e-03 & 2.775e-03 & 5.265e-12 & 3.068e-12 & 2.521e-12 & 2.462e-03 \\
        10 & 2.258e-03 & 2.775e-03 & 5.763e-13 & 4.402e-13 & 3.411e-13 & 2.462e-03 \\
        25 & 2.258e-03 & 2.775e-03 & 5.756e-12 & 3.421e-12 & 3.565e-12 & 2.462e-03 \\
        50 & 2.258e-03 & 2.775e-03 & 4.369e-12 & 2.318e-12 & 2.752e-12 & 2.462e-03 \\
        \hline
      \end{tabular}
    \end{table}

    \begin{table}[tb]
      \caption{Final time steps and cost saving factors computed with Eq.  \eqref{eqn:CostSavings}, obtained by the step size control algorithm (Algorithm
               \ref{alg:StepSizeControl}) using different values of $n_s$ and the reference parameter vectors taken from Table \ref{table:ParameterValues-Modelhierarchy}.}
      \label{table:StepSizeControlStepCostSaving}
      \centering
      \begin{tabular}{c c c c c c c}
        \hline
         & N & N-DOP & NP-DOP & NPZ-DOP & NPZD-DOP & MITgcm-PO4-DOP \\
        \\hline
         final time step & 32$\Delta t$ & 32$\Delta t$ &1$\Delta t$ &1$\Delta t$&1$\Delta t$ &32$\Delta t$ \\
         cost saving for $n_{s}$ = 1 &96 & 96 & -0.5 & -0.5 & -0.5 & 96 \\
         for $n_{s}$ = 2& 95& 95 & -0.5 & -0.5 & -0.5 &95\\
         for $n_{s}$ = 5&92 & 92 & -0.5 & -0.5 & -0.5 &92\\
         for $n_{s}$ = 10&88 & 88 & -0.5 & -0.5 & -0.5 &88\\
         for $n_{s}$ = 25&74 &74  & -0.5 & -0.5 & -0.5 &74\\
         for $n_{s}$ = 50&51 & 51 & -0.5 & -0.5 & -0.5 &51\\
           \hline
      \end{tabular}
    \end{table}

    The number of model years $n_s \in \mathbb{N}$, after which the error
    estimation was computed, had no influence on the accuracy of the steady
    annual cycle approximation using the step size control (Table
    \ref{table:StepSizeControlStepYear}). For the N, N-DOP and MITgcm-PO4-DOP 
    models, the step size control increased the time step as quickly as possible
    to the maximal time step regardless of the number of model years $n_s$ and,
    afterwards, applied only this maximal time step. As a consequence, the
    approximations for the different model years $n_s$ were almost identical.
    Similar to the results above for the NP-DOP, NPZ-DOP and NPZD-DOP models,
    the approximation corresponded to the reference solution for the different
    model years $n_s$ since the step size control used always time step
    1\,$\Delta t$. As can be seen in Table \ref{table:StepSizeControlStepCostSaving}, the total cost saving depends on the parameter $n_{s}$. Taking a bigger value, the algorithm needs more model years to reach the final, maximal step size. Thus, the cost saving decreases with growing values of $n_{s}$.
   
    \begin{table}[tb]
      \caption{Relative error \eqref{eqn:relativeError} for $\ell = 10000$
               obtained by the step size control algorithm (Algorithm
               \ref{alg:StepSizeControl}) using different norms for the local
               error estimation.}
      \label{table:StepSizeControlNorms}
      \centering
      \begin{tabular}{l c c c c c c}
        \hline
        Norm & N & N-DOP & NP-DOP & NPZ-DOP & NPZD-DOP & MITgcm-PO4-DOP \\
        \\hline
        $\left\| \cdot \right\|_2$ & 2.258e-03 & 1.214e-03 & 3.439e-12 & 1.744e-12 & 1.845e-12 & 2.462e-03 \\
        $\left\| \cdot \right\|_{2,V}$ & 2.258e-03 & 2.775e-03 & 3.439e-12 & 1.744e-12 & 1.845e-12 & 2.462e-03 \\
        $\left\| \cdot \right\|_{2, T}$ & 2.258e-03 & 2.775e-03 & 5.597e-12 & 5.537e-04 & 4.738e-12 & 2.462e-03 \\
        $\left\| \cdot \right\|_{2, V, T}$ & 2.258e-03 & 2.775e-03 & 6.213e-03 & 1.383e-03 &     1.398e-03     & 2.462e-03 \\
        $\left\| \cdot \right\|_{2, V}|_{\{1\}}$ & 1.215e-12 & 1.885e-12 & 1.849e-12 & 6.879e-13 & 1.005e-12 & 2.195e-12 \\
        $\left\| \cdot \right\|_{2, V}|_{\{1, 2, 3\}}$ & 5.021e-04 & 1.885e-12 & 1.849e-12 & 6.879e-13 & 1.005e-12 & 2.195e-12 \\
        $\left\| \cdot \right\|_{2, V}|_{\{11\}}$ & 2.258e-03 & 2.775e-03 & 1.355e-02 & 1.804e-02 & 1.911e-02 & 2.462e-03 \\
        \hline
      \end{tabular}
    \end{table}

    The norm used to estimate the local discretization error affected the
    approximation calculated with the step size control (Table
    \ref{table:StepSizeControlNorms}) in some cases. Firstly, the application of
    the Euclidean norm $\left\| \cdot \right\|_2$ instead of the norm $\left\|
    \cdot \right\|_{2, V}$ influenced the step size control for the N-DOP model
    only. For this model, the step size control, occasionally, did not increase
    the time step directly to the maximal time step and finished the simulation
    with time step 16\,$\Delta t$. This led to a smaller relative error.
    The use of the norms $\left\| \cdot \right\|_{2,T}$ and $\left\| \cdot
    \right\|_{2,V,T}$, secondly, resulted in at least one increase of the time
    step for the NP-DOP, NPZ-DOP and NPZD-DOP models and, therefore, in a
    reduction of the computational costs. However, these norms had no influence
    for the N, N-DOP and MITgcm-PO4-DOP models. Lastly, the restriction of the
    norm $\left\| \cdot \right\|_{2,V}$ to different layers had an impact on the
    error estimation because the annual variability exclusively affects
    concentrations in the upper ocean, while the concentrations in the deeper
    ocean change very slowly. Besides, the box volumes in the upper ocean are
    small in comparison to the box volumes in the deeper ocean, but big boxes
    have a greater effect on the norm as small ones for a volume weighted norm.
    However, the local discretization error was high in the upper ocean due to
    the variability, and small in the deeper ocean. Thus, the step size control
    increased more frequently the step size using the restricted norms including
    deeper layers or including more layers (Table
    \ref{table:StepSizeControlNorms}).
  
    \begin{table}
      \caption{Relative error \eqref{eqn:relativeError} for $\ell = 10000$
               obtained by the step size control algorithm (Algorithm
               \ref{alg:StepSizeControl}) using different initial time steps.}
      \label{table:StepSizeControlInitialTimeStep}
      \centering
      \begin{tabular}{c c c c c c c}
        \hline
        $m_{\text{init}}$ & N & N-DOP & NP-DOP & NPZ-DOP & NPZD-DOP & MITgcm-PO4-DOP \\
        \\hline
         1 & 2.258e-03 & 2.775e-03 & 3.439e-12 & 1.744e-12 & 1.845e-12 & 2.462e-03 \\
         2, 4, 8, 16, 32 & 2.258e-03 & 2.775e-03 & 1.848e-12 & 5.537e-04 & 7.119e-10 & 2.462e-03 \\
        \hline
      \end{tabular}
    \end{table}

    The choice of the initial time step especially affected the three most
    complex biogeochemical models. Due to the directly increasing time step up
    to the maximum, no differences were visible for the N, N-DOP and
    MITgcm-PO4-DOP models (Table \ref{table:StepSizeControlInitialTimeStep}). In
    contrast, the relative errors in Table
    \ref{table:StepSizeControlInitialTimeStep} indicate the use of larger time
    steps for at least one model year for the other biogeochemical models. In
    particular, the step size control utilized time step 2\,$\Delta t$ over the
    entire simulation for initialization with larger time steps for the NPZ-DOP
    model.

    \begin{table}[tb]
      \caption{Relative error \eqref{eqn:relativeError} for $\ell = 10000$
               obtained by the step size control algorithm (Algorithm
               \ref{alg:StepSizeControl}) using different tolerances $\tau_0
               \in \{0.1, 0.2, \ldots, 1.0\}$.}
      \label{table:StepSizeControlTolerance}
      \centering
      \begin{tabular}{c c c c c c c}
        \hline
        $\tau_0$ & N & N-DOP & NP-DOP & NPZ-DOP & NPZD-DOP & MITgcm-PO4-DOP \\
        \\hline
        1.0, 0.9, \ldots, 0.5 & 2.258e-03 & 2.775e-03 & 3.439e-12 & 1.744e-12 & 1.845e-12 & 2.462e-03 \\
        0.4 & 2.258e-03 & 5.106e-04 & 3.439e-12 & 1.744e-12 & 1.845e-12 & 2.462e-03 \\
        0.3 & 2.258e-03 & 3.794e-12 & 3.439e-12 & 1.744e-12 & 1.845e-12 & 4.872e-04 \\
        0.2, 0.1 & 2.171e-12 & 3.794e-12 & 3.439e-12 & 1.744e-12 & 1.845e-12 & 3.277e-12 \\
        \hline
      \end{tabular}
    \end{table}

    The tolerance $\tau_0 \in \mathbb{R}_{>0}$ influenced only the time step's
    increase for the N, N-DOP and MITgcm-PO4-DOP models. Using smaller
    tolerances, the possible increase to the maximal time step required more
    model years for the N, N-DOP and MITgcm-PO4-DOP models. The use of the
    smallest tolerance always resulted in time step 1\,$\Delta t$ for the
    entire spin-up. For the N-DOP and MITgcm-PO4-DOP models, there existed a
    tolerance for which the step size control increased the time step only up to
    8\,$\Delta t$ (Table \ref{table:StepSizeControlTolerance}). For the NP-DOP,
    NPZ-DOP and NPZD-DOP models, the algorithm did not increase the step when
    tolerance $\tau_0 = 1.0$ was used. Consequently, the choice of a smaller
    tolerance had no effect (Table \ref{table:StepSizeControlTolerance}).

\subsection{Algorithm \ref{alg:DecreasingTimesteps}: Decreasing Time Steps}
\label{sec:Results-DecreasingTimesteps}

  \begin{figure}[tbp]
    \centering
    \includegraphics{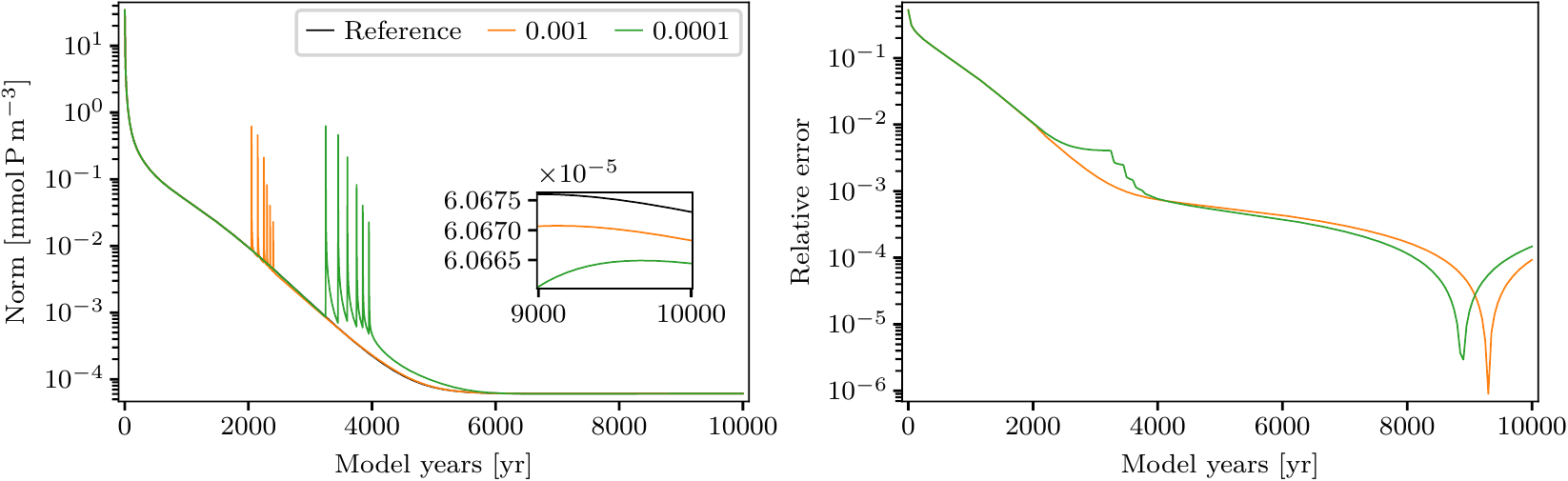}
    \caption{Results of spin-up computations for the reference parameter vector
             (see Table \ref{table:ParameterValues-Modelhierarchy}) and the N
             model using the decreasing time steps algorithm (Algorithm
             \ref{alg:DecreasingTimesteps}). Shown are the convergence of the
             spin-up (left), i.e., the norms of difference
             \eqref{eqn:StoppingCriterion} between consecutive iterations, and
             the relative error \eqref{eqn:relativeError} (right) using different
             tolerances $\varepsilon \in \{0.001, 0.0001\}$. Furthermore, the
             figure of the norm of difference (left) contains the convergence
             towards a steady annual cycle for the reference solution. In this
             figure, it is mostly covered by that of the decreasing time steps
             algorithm with tolerance $\varepsilon = 0.001$.}
    \label{fig:DecreasingTimesteps_Convergence}
  \end{figure}

  The decreasing time steps algorithm computed a reasonable approximation of
  the steady annual cycle with the automatic stepwise reduction of the time
  step and, thus, reduced the computational effort. For the N model with the
  parameter vector listed in Table \ref{table:ParameterValues-Modelhierarchy},
  Fig. \ref{fig:DecreasingTimesteps_Convergence} demonstrates the similar
  convergence behavior towards a steady annual cycle using the decreasing time
  steps algorithm as well as the spin-up calculation of the reference
  solution. Using the decreasing time steps algorithm, the six peaks in the
  norm of differences \eqref{eqn:StoppingCriterion} pertained to the six
  reductions of the time step. As a result of the decreased time step, a large
  concentration change took place in one model year similar to the large
  changes at the beginning of each spin-up. In particular, the reduction of
  the relative error resulting from the decrease of the time step is evident
  in Fig. \ref{fig:DecreasingTimesteps_Convergence} (right) using the
  tolerance 0.0001. Although the algorithm applied time step 1\,$\Delta t$ at
  the end of the spin-up computation, the spin-up did not perfectly converge
  against the reference solution. Nevertheless, the algorithm shortened the
  runtime of the simulation, and finished with an approximation of the steady
  annual cycle that was much better than the one of the spin-up with constant
  time step 64\,$\Delta t$, i.e., $\mathbf{y}^{10000, 64}$ (cf. Fig.
  \ref{fig:Timesteps_Convergence}). The used tolerance affected only slightly
  the accuracy of the approximation, but the use of a smaller tolerance resulted
  in a cost saving because the decreasing time steps algorithm used each time
  step longer before the time step was reduced. Namely, the application of the
  decreasing time steps algorithm with tolerance 0.001 resulted in a cost
  saving of 23\% compared to the spin-up computation of the reference solution
  and in a cost saving of 38\% using the tolerance 0.0001. The results for the
  other biogeochemical models using the reference parameter vectors are
  parallel to the results shown for the N model.

  \begin{figure}[!tb]
    \centering
    \includegraphics{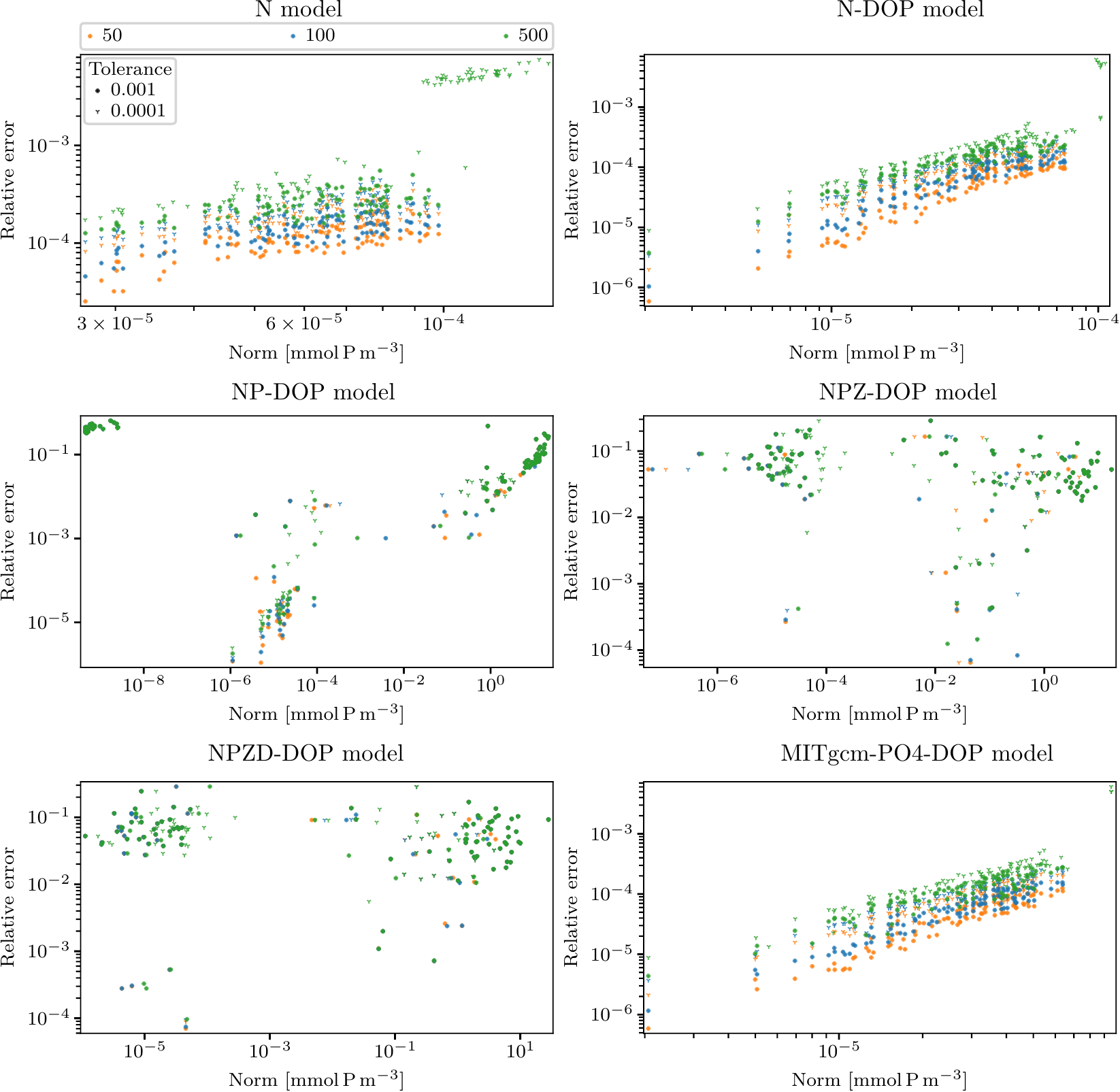}
    \caption{Norm of difference \eqref{eqn:StoppingCriterion} and relative error
             \eqref{eqn:relativeError} for $\ell = 10000$ using different
             configurations (i.e., tolerance $\varepsilon \in \{0.001,
             0.0001\}$ and number of model years $n_s \in \{50, 100, 500\}$)
             of the decreasing time steps algorithm. Shown are the results for
             the different biogeochemical models and all parameter vectors of
             the Latin hypercube sample.}
    \label{fig:DecreasingTimesteps_Scatter}
  \end{figure}

  \begin{figure}[!tb]
    \centering
    \includegraphics{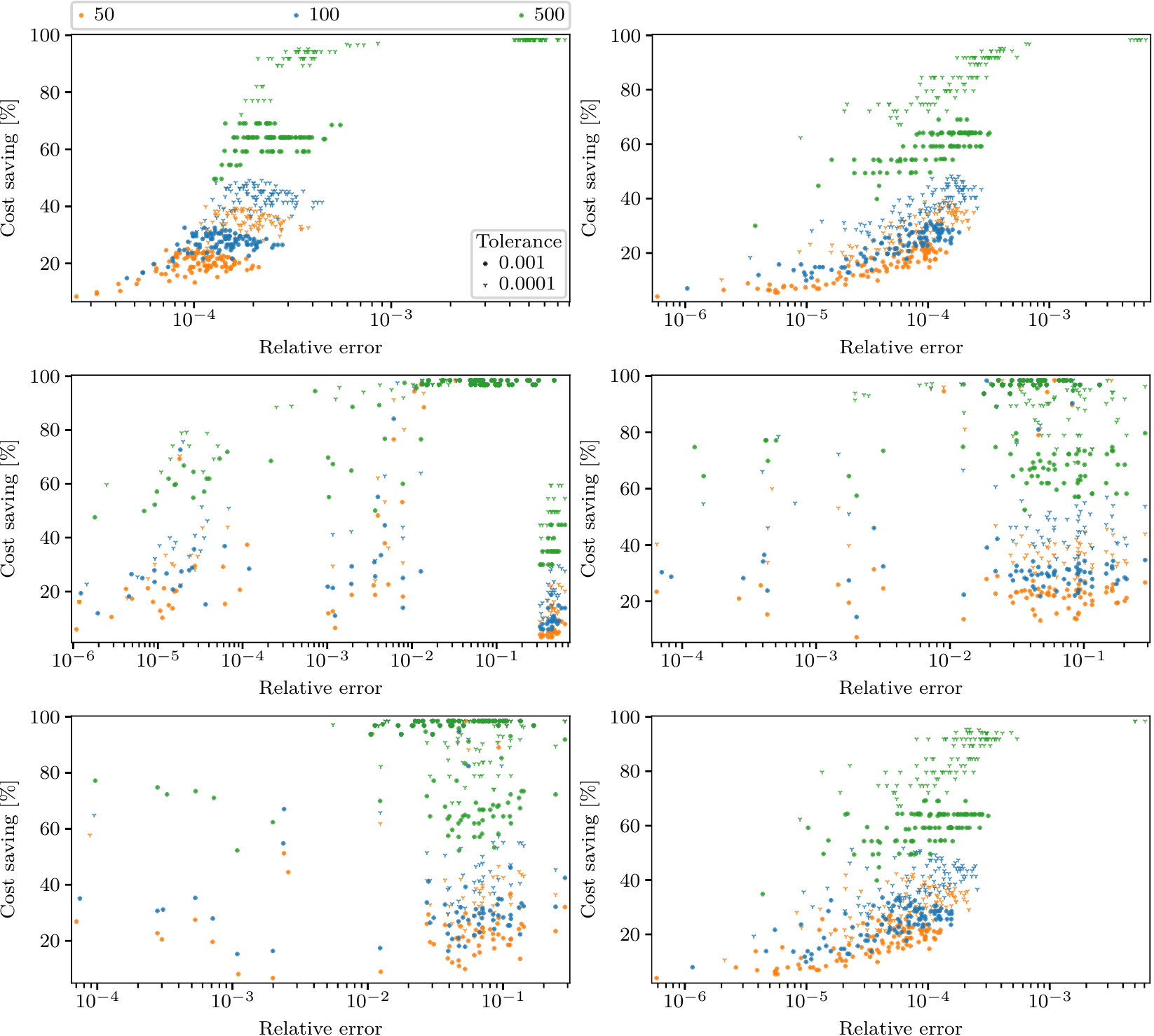}
    \caption{Saving of computational costs \eqref{eqn:CostSavings}
             using the decreasing time steps algorithm with different
             configurations (i.e., tolerance $\varepsilon \in \{0.001,
             0.0001\}$ and number of model years $n_s \in \{50, 100, 500\}$).
             Shown are the results for the different biogeochemical models and
             all parameter vectors of the Latin hypercube sample (top left: N
             model, top right: N-DOP model, middle left: NP-DOP model, middle
             right: NPZ-DOP model, bottom left: NPZD-DOP model, bottom right:
             MITgcm-PO4-DOP model).}
    \label{fig:DecreasingTimesteps_CostSaving}
  \end{figure}

  The decreasing time steps algorithm avoided a divergent spin-up calculation
  but did not decrease the time step to the minimum for each simulation. We
  analyzed the decreasing time steps algorithm using the parameter vectors of
  the Latin hypcercube sample for the different biogeochemical models. For the
  N, N-DOP and MITgcm-PO4-DOP models, the decreasing time steps algorithm
  computed an appropriate approximation of the steady annual cycle for the
  different configurations of the tolerance $\varepsilon \in \{0.001,
  0.0001\}$ and the number of model years $n_s \in\{50, 100, 500\}$ as
  detailed in Fig. \ref{fig:DecreasingTimesteps_Scatter} (cf. Fig.
  \ref{fig:Timesteps_Convergence}). More specifically, the accuracy
  decreased slightly, on the one hand, with a smaller tolerance and, on the
  other hand, with a larger number of model years $n_s$ because in both cases
  larger time steps were used over a longer period of model years. Figure
  \ref{fig:DecreasingTimesteps_CostSaving} reflected this behavior in the cost
  saving. The computational effort increased as a result of a more frequent
  check of the time step reduction or with a smaller tolerance range because
  the time step tended to be reduced earlier. For the NP-DOP, NPZ-DOP and
  NPZD-DOP models, the approximations calculated with the decreasing time steps
  algorithm were often identical for the different configurations. The algorithm decreased  the time step to
  1\,$\Delta t$  only in
  half of the simulation runs, while this occurred in more than 90\% of the simulations
  using the other three biogeochemical models. In fact, the criterion checking
  the decrease of the time step was inappropriate if the norm of differences
  \eqref{eqn:StoppingCriterion} oscillated, cf. \cite{PfeSla21a}. The
  algorithm, therefore, applied large time steps during  the entire spin-up, wherefore
  the relative error was high (Figs. \ref{fig:DecreasingTimesteps_Scatter} and
  \ref{fig:Timesteps_Convergence}). In some of these cases, the algorithm temporarily decreased the time step.
However, this hardly effected the accuracy of the approximation
  (Fig. \ref{fig:DecreasingTimesteps_CostSaving}). However, the decreasing time
  steps algorithm automatically reduced the time step if the simulation
  diverged due to a too large time step. In contrast to the simulations with a
  low accuracy of the approximation using the NP-DOP, NPZ-DOP and NPZD-DOP
  models, there were also parameter vectors for which the decreasing time steps
  algorithm calculated a reasonable approximation with reduced computational
  costs (Fig. \ref{fig:DecreasingTimesteps_CostSaving}).

%% file: Latexfiles/Conclusions_arxiv.tex
We presented and evaluated three algorithms that 
automatically adjust the temporal step size in the spin-up of marine ecosystem models.  
The setting we used was an offline computation where the ocean transport is realized by transport matrices.
We tested the algorithms for six biogeochemical models with different complexity. 
The first two algorithms are based on a mathematical estimate of the local error taken over a given number of model years. The algorithms are able to both increase and decrease the step size during the spin-up, depending on this error estimate. The  first algorithm ignores any negative tracer concentrations, whereas the second adjusts the step size such that negative values are avoided. The third algorithm starts with a coarse step size and reduces it along the convergence of the spin-up. It does not use any mathematical error estimate.

First  of all, all algorithms properly worked in the intended way for  the investigated model configurations. The spin-up calculations yielded approximations of the
steady annual cycle which were in agreement with the
respective reference solution obtained with the standard time step. This holds
for the considered hierarchy of six biogeochemical models and, for each model,
for a reference parameter vector as well as for a Latin hypercube sample of 100
parameter vectors. 

We now discuss the behavior of all three algorithm and draw conclusions for them separately.  Algorithm 1 automatically adjusts the time step based on the error estimate, but does not take into account any negative tracer values that occur. The error estimation itself needs an additional time-step and, thus, creates an overhead. For the three simpler biogeochemical models, this algorithm  chose the biggest feasible step size directly at the beginning of the spin-up. As a consequence, the reduction of the computational effort was very high (up to 95\%), since at the upper limit of the step size no further error estimation was performed, and the mentioned overhead did not occur. 
In contrast, for the three biogeochemical models with higher complexity the algorithm always used the smallest step size. Hence, no performance gain due to bigger time-steps was realized, and, even worse, the error estimation caused maximal overhead. There are some options to vary the behavior of the algorithm. One is the length of the error estimation period. In our computations, this made no qualitative difference with respect to the chosen step size, but only to the overall performance gain.
Two other options that can be easily realized, but were not investigated here, are to estimate the error either only in the beginning  or in other selected temporal sub-intervals of the spin-up, but not all the time. Both options will definitely reduce the overhead.  
However, an adaptive step size control is designed to react on different behavior in different time intervals. Thus, a restriction of  the error estimation on certain time pre-defined sub-intervals in the spin-up is somehow arbitrary. It would require some prior knowledge or assumption of the convergence behavior during the spin-up. 
Another option would be to apply an adaptive time-stepping \textit{within} one year, and use this choice throughout the whole spin-up. This will also reduce the overhead, since the period for error estimation would be drastically shortened.

Algorithm 2 which additionally strictly avoids negative tracer values did not lead to any performance gain. In all our experiments, it always selected the smallest feasible time-step. Again the error estimation led to a huge overhead. This result  shows that these kind of spin-up calculations  rely on the acceptance of (small) negative values, even though, in theory, the solutions should stay non-negative for all times. It can be seen that the spin-up results are reasonable even with  small negative values occurring at some spatial points.

Algorithm 3 which decreased the time-step during the spin-up depending on its convergence to the steady annually periodic state does not suffer from any computational overhead. It reduces the runtime for the spin-up significantly. The actual reduction depends on the chosen model and its parameters.

As a summary, we can only recommend to use Algorithm 3 directly for an actual spin-up computation. This algorithm is also the easiest to realize. 
The failing of Algorithm 2 with respect to runtime reduction rises the question how important local negative tracer values are at all for the spin-up runs. This is a question that is beyond the scope of this paper. 
Finally, from the results for the  step size control 
with Algorithm 1 we deduce that the simpler models do not show different temporal behavior at all during the spin-up, and the more complex do not allow to decrease the temporal resolution. These are results about the models and their behavior in a spin-up: For the simpler models, this means that their spin-up convergence is rather uniform. 
However, for other existing far more complex models (see, e.g., \cite{ISSMLN13}) the question remains if these models also show this uniform spin-up convergence. For such kind of models, Algorithm 1 might be a useful tool to study the convergence behavior during the spin-up.

%% file: Latexfiles/CodeDataAvailability_arxiv.tex
The code used to generate the data in this publication is available at
\url{https://doi.org/10.5281/zenodo.6397419} and \url{https://metos3d.github.io} \cite{PfeSla21cCode}.
All used and generated data are available at
\url{https://doi.org/10.5281/zenodo.5644003} \cite{PfeSla21cData}.